\documentclass[times,twocolumn,final,authoryear]{elsarticle}

\usepackage{ycviu}
\usepackage{framed,multirow}

\usepackage{amssymb}
\usepackage{latexsym}

\usepackage{url}
\usepackage{xcolor}

\usepackage{graphicx}
\graphicspath{{./figures/}}

\usepackage{amssymb,amsfonts,amsmath,amscd}

\usepackage{bm} 

\usepackage{lipsum}

\usepackage[all]{nowidow}

\usepackage{algpseudocode}
\PassOptionsToPackage{hyphens}{url}
\usepackage{url,hyperref,lineno,microtype,subcaption}
\usepackage{booktabs}%
\usepackage{tabu}%
\usepackage{longtable}%
\usepackage{rotating,tabularx}
\usepackage{multirow}
\usepackage{threeparttable}

\usepackage{makecell}

\usepackage{siunitx}
\usepackage[printonlyused]{acronym}

\acrodef{BAP}{blur aware plenoptic}
\acrodef{MFPC}{multi-focus plenoptic camera}
\acrodef{MLA}{micro-lenses array}
\acrodef{MIA}{micro-images array}
\acrodef{MI}{micro-image}
\acrodef{MIC}{micro-image center}
\acrodef{SAI}{sub-aperture image}
\acrodef{CoC}{circle of confusion}
\acrodef{DoF}{depth of field}
\acrodef{EPI}{epipolar plane image}
\acrodef{CSAD}{central sub-aperture depth map}

\acrodef{PnP}{Perspective-n-Point}
\acrodef{SfM}{Structure-from-Motion}
\acrodef{PIV}{particle image velocimetry}
\acrodef{SPO}{spinning parallelogram operator}
\acrodef{MRF}{Markov random field}
\acrodef{CNN}{convolutional neural network}

\acrodef{RMSE}{root-mean-square error}
\acrodef{MBE}{mean bias error}
\acrodef{PSF}{point-spread function}

\newcommand{\ie}{i.e.,~}
\newcommand{\eg}{e.g.,~}

\newcommand{\tcomma}{\text{, }}
\newcommand{\tdot}{\text{.}}

\newcommand{\Field}[1]{\mathbb{#1}}

\newcommand{\Group}[1]{\mathrm{#1}}

\newcommand{\Reals}{\Field{R}}

\newcommand{\SE}[1]{\Group{SE}\left(#1\right)}

\newcommand{\vect}[1]{\bm{#1}}
\newcommand{\mat}[1]{\bm{#1}}
\newcommand{\bbm}{\begin{bmatrix}}
	\newcommand{\ebm}{\end{bmatrix}}

\newcommand{\T}{^\top}%

\newcommand{\fun}[1]{\mathrm{#1}}
\newcommand{\funarg}[2]{\fun{#1}\!\left(#2\right)}

\newcommand{\Proj}{\fun{\Pi}}

\newcommand{\Neighb}{\mathcal{N}}
\newcommand{\neighb}[1]{\Neighb\left(#1\right)}

\newcommand{\set}[1]{\left\lbrace#1\right\rbrace}
\newcommand{\range}[1]{\left[#1\right]}
\newcommand{\norm}[1]{\left\Vert#1\right\Vert}
\newcommand{\abs}[1]{\left\vert#1\right\vert}

\newcommand{\convsym}{\text{\,\textasteriskcentered~}}
\newcommand{\convp}[2]{#1\convsym#2}

\newcommand{\nump}[2]{\num[round-mode=places,round-precision=#1]{#2}}

\newcommand{\focal}{f}
\newcommand{\Focal}{F}

\newcommand{\baseline}{B}

\newcommand{\focusdist}{h}

\newcommand{\virtdepth}{\upsilon}%

\newcommand{\disparity}{\delta}

\newcommand{\blurradiusmm}{r}
\newcommand{\blurradiuspix}{\rho}

\newcommand{\distmlasensor}{\dist}
\newcommand{\distlensmla}{\Dist}

\newcommand{\distobj}{a}

\newcommand{\mlinterdist}{\Delta\!{\mlcenter}}%
\newcommand{\mlcenter}{\vect{C}}
\newcommand{\mlindexes}{k,l}

\newcommand{\micenter}{\vect{c}}

\newcommand{\principalpointx}{u_0}
\newcommand{\principalpointy}{v_0}

\newcommand{\pixelsize}{s}%

\newcommand{\Distor}{\fun{\varphi}}
\newcommand{\distor}[1]{\funarg{\varphi}{#1}}

\newcommand{\invdistor}[1]{\funarg{\varphi^{-1}}{#1}}

\newcommand{\Psf}{\funarg{h}{x,y}}
\newcommand{\psf}{\fun{h}}

\newcommand{\Kthinlens}{\mat{K}\!}

\newcommand{\Transform}{\mat{T}}

\newcommand{\poseMLkl}{\Transform\fmla\!\left(\mlindexes\right)}
\newcommand{\poseLens}{\Transform\fcam}

\newcommand{\bapprojm}{\mat{\mathcal{P}}\!} %

\newcommand{\invKthinlens}{\mat{K}{}^{-1}\!}
\newcommand{\invposeMLkl}{\Transform{}^{-1}\fmla\!\left(\mlindexes\right)}
\newcommand{\invposeLens}{\Transform{}^{-1}\fcam}
\newcommand{\invbapprojm}{\mat{\mathcal{P}}{}^{-1}\!} %

\newcommand{\Frame}[1]{\mathcal{#1}}

\newcommand{\fworld}{_w}
\newcommand{\fcam}{_c}
\newcommand{\fmla}{_\mu}

\newcommand{\fundistorted}{_u}
\newcommand{\fdistorted}{_d}

\newcommand{\mltype}[1]{{}^{\left(#1\right)}}
\newcommand{\mltypenb}{I}

\newcommand{\Img}{\mathcal{I}}
\newcommand{\img}[1]{\Img\!\left(#1\right)}

\newcommand{\point}{\vect{p}}

\newcommand{\dist}{d}
\newcommand{\Dist}{D}

\newcommand{\feature}{\vect{p}}

\journal{Computer Vision and Image Understanding}

\begin{document}

\ifpreprint
  \setcounter{page}{1}
\else
  \setcounter{page}{1}
\fi

\begin{frontmatter}

\title{Blur aware metric depth estimation with multi-focus plenoptic cameras}

\author[1]{Mathieu \snm{Labussi\`{e}re}\corref{cor1}} 
\ead{mathieu.labussiere@uca.fr}
\author[1]{C\'{e}line \snm{Teuli\`{e}re}}
\author[1]{Omar \snm{Ait-Aider}}

\address[1]{Universit\'{e} Clermont Auvergne, Clermont Auvergne INP, CNRS, Institut Pascal, F-63000 Clermont-Ferrand, France}
\cortext[cor1]{Corresponding author:}

\received{}
\finalform{}
\accepted{}
\availableonline{}
\communicated{}

\begin{abstract}
While a traditional camera only captures one point of view of a scene, a plenoptic or light-field camera, is able to capture spatial and angular information in a single snapshot, enabling depth estimation from a single acquisition.
In this paper, we present a new metric depth estimation algorithm using only raw images from a {multi-focus} plenoptic camera.
The proposed approach is especially suited for the multi-focus configuration where several micro-lenses with different focal lengths are used.
The main goal of our blur aware depth estimation (BLADE) approach is to improve disparity estimation for defocus stereo images by integrating both correspondence and defocus cues.
We thus leverage blur information where it was previously considered a drawback.
We explicitly derive an inverse projection model including the defocus blur providing depth estimates up to a scale factor. 
A method to calibrate the inverse model is then proposed.
We thus take into account depth scaling to achieve precise and accurate metric depth estimates.
Our results show that introducing defocus cues improves the depth estimation.
We demonstrate the effectiveness of our framework and depth scaling calibration on relative depth estimation setups and on real-world 3D complex scenes with ground truth acquired with a 3D lidar scanner. 
\end{abstract}


\end{frontmatter}

\section{Introduction}\label{sec:introduction}
\begin{figure}[!t]
\centering
\includegraphics[]{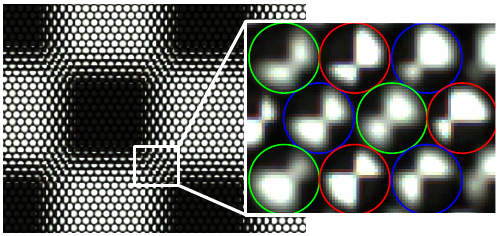}
\caption{
	Example of raw plenoptic image multiplexing both angular and spatial information onto the sensor in the form of a Micro-Images Array (MIA) with several types of micro-lenses, thus different amounts of blur.
}\label{fig:rawimage}
\end{figure}

Plenoptic or light-field cameras are imaging systems able to capture both \textit{spatial} and \textit{angular} information about a scene in a single exposure.
These systems are usually built upon a \acf{MLA} placed between a main lens and a sensor \citep{Ng2005b,Perwass2010c,Georgiev2012}.
Information from light is multiplexed onto the sensor in the form of a \ac{MIA} as illustrated in \autoref{fig:rawimage}.
With \textit{unfocused} plenoptic cameras (1.0), the micro-lenses are focused at infinity, and each pixel captures a specific orientation.
With \textit{focused} plenoptic cameras (2.0), the imaging process can be modeled as the projection of object points by the main lens into a virtual intermediate space which are then re-imaged by each micro-lens onto the sensor.
To improve the spatial resolution and \ac{DoF} of the plenoptic camera, a multi-focus configuration has been proposed by \cite{Perwass2010c,Georgiev2012}.
In this setup, the \ac{MLA} is composed of several micro-lenses with different focal lengths.
The same part of a scene is projected into multiple observations on the sensor with different amounts of blur according to the micro-lens' type, as shown in the zoom of \autoref{fig:rawimage}.
{In this paper, we especially focus on this latter setup, i.e., multi-focus plenoptic cameras.}
Plenoptic or light-field data can designate different quantities: either it is referring to the multi-views array of reconstructed \acp{SAI} as by \cite{Wu2017}, or to the micro-images array from raw images.
The term \textit{view} is applied either to a \ac{SAI} or to a micro-image, depending on the context.

From this redundant information, depth estimation and 3D reconstruction can be performed directly from a single acquisition, with scale information.
First, plenoptic or light-field depth estimation can be seen as a standard multi-view stereo matching problem.
Depth can be inferred for a reference view, as in \textit{Depth from Stereo} approaches \citep{Scharstein2002}, using \textit{correspondence} cues.
Second, as the plenoptic camera captures part of the scene at different focus settings (either using refocused \acp{SAI} or different micro-images' type), \textit{defocus} cues can be used to estimate depth like in \textit{Depth from Focus/Defocus} approaches \citep{Pentland1987,Grossmann1987,Subbarao1989,Lai1992}.

In the focused plenoptic camera case, working with \acp{SAI} is prone to error as depth is usually required to reconstruct the light-field or the \acp{SAI} {\citep{Hog2017b,Liang2015,Wanner2011b}}.
To overcome this issue, algorithms can work directly with the raw plenoptic images, at micro-images level.
However, usually only micro-images with the smallest amount of blur are used, or alternatively, specific patterns are designed to exploit the information \citep{Fleischmann2014,Ferreira2016,Palmieri2017}.
Using the model of \cite{Labussiere2020blur} extended by \cite{Labussiere2022calib}, we relate the camera parameters to the amount of blur in the image, and all information can be used simultaneously, without distinction between types of micro-lenses.
We propose then to leverage blur information where it was previously considered a drawback.
Indeed defocus cues are complementary to correspondence cues, and can improve the quality of depth estimation as shown by \cite{Tao2013}.\\

\noindent\textbf{Contributions.}
As our main contribution, we introduce in this paper a metric depth estimation framework for plenoptic cameras, named blur aware depth estimation (BLADE), leveraging both spatially-variant blur and disparity cues between micro-images, using only raw images from plenoptic cameras.
It is especially suited for the multi-focus configuration where several micro-lenses with different focal lengths are used.
The method uses the camera model presented in \cite{Labussiere2022calib} which explicitly includes the defocus blur in the projection model.
{We derive its inverse which provides depth estimates up to a scale factor. Thus, we {empirically} take into account depth scaling to achieve precise and accurate metric depth estimates. 
For that purpose, calibration is required. Especially, we propose a method to calibrate the inverse model.%
}
To validate our framework, we present a new dataset of 3D real-world scenes with ground truths acquired with a 3D lidar scanner, and a methodology to calibrate the extrinsic parameters. 
We make our source code and datasets publicly available to the community {on our github page's \url{https://github.com/comsee-research/}}. \\

\noindent\textbf{Paper Organization.}
The remainder of this paper is organized as follows.
First, we review the existing methods for depth estimation based on light-field data in \autoref{sec:rw}.
Second, we briefly recall the blur aware plenoptic (BAP) camera model and present the {derivation of the} inverse projection model in \autoref{sec:cameramodel}{, as well as, how depth scaling is taken into account and calibrated in \autoref{sec:depthcalibration}.}
Third, we explain in \autoref{sec:linkdefocusdisp} how we can link the disparity in image space to the defocus blur information.
Then, we detail our blur aware depth estimation (BLADE) framework in \autoref{sec:blade}. %
Our experimental setup is presented in \autoref{sec:setup}.
Finally, our results are presented and discussed in \autoref{sec:res}. 

\noindent\textbf{Notations.} 
The camera notations used in this paper are shown in \autoref{fig:notations}.
Pixel counterparts of metric values are denoted in lower-case Greek letters.
Bold font represents vectors (usually in lower-case letters) and matrices ({usually} in upper-case letters). 
Scalars are given by light letters.

\section{Related work}\label{sec:rw}
First, we briefly review classical paradigms for depth estimation and how they apply to plenoptic imaging.
Then, we extensively analyze the different approaches to estimate depth from light-field.

\subsection{Depth from Stereo}

Depth from stereo is achieved by triangulating rays passing through corresponding regions or features from two images. 
Correspondences are generally obtained based on local photometric similarity measurement. 
The global consistency is also achieved by taking into account epipolar geometry constraint. 
To make the process simpler and faster, the stereo pair can be rectified in order to handle non-parallel image planes \citep{Scharstein2002}.
In the case of plenoptic cameras, micro-images share the same plane. 
Thus, there is no need for rectification.
However, in the multi-focus case, the similarity constraint is violated due to the difference of focus between two micro-images types.
To satisfy the constraint, one might only compare images of the same type, or restrict the working range to the \ac{DoF} of the camera (as at least two micro-images are in-focus simultaneously).
We propose an alternative approach that exploits all the available information by including blur within the depth estimation process.

\subsection{Depth from Focus/Defocus}

The depth from focus/defocus approach aims to estimate the spatially variant spread parameter of the blur kernel, by acquiring two images of the same scene with different camera settings \citep{Pentland1987,Grossmann1987,Subbarao1989,Lai1992}.
The blur radius is linked to the inverse distance, and once the blur is estimated, depth can be retrieved.
The spread parameter is usually estimated in the frequency domain or in the spatial domain \citep{Subbarao1994}.
Depth from focus/defocus works better on short ranges and can thus be seen as a solution for these distances.
Indeed, depth from stereo methods are less effective in a short range due to part of the scene being not visible by both cameras.
However, the main hypothesis is that the two images represent the same scene viewed from the same point of view and with the same view angle, which is not the case with a plenoptic camera. 
Therefore, %
the micro-images have to be matched according to parallax.

\subsection{Depth from Light-Field}

Most light-field depth estimation processes are divided into two steps: 1) initial depth map estimation from \acp{SAI} or \acp{EPI}, and 2) depth refinement with global methods.
These kinds of methods usually operate with \textit{unfocused} plenoptic cameras (1.0).

\subsubsection{Based on sub-aperture images (SAIs)}

One category of approaches estimates depth from reconstructed \acp{SAI}.
A framework is proposed by \citet{Nava2009} to simultaneously estimate an {all in-focus} image along with the depth map, based on focal stack analysis for the focused plenoptic camera.
\citet{Bishop2012} applied traditional multi-view stereo methods along with an explicit image formation model and antialiasing to reconstruct both scene depth and its super-resolved texture in a Bayesian framework. 
In \cite{Hahne2018b}, light field triangulation has been proposed to determine depth distances and baselines in unfocused plenoptic cameras. However their model is not applicable to the focused plenoptic camera.
Correspondence and defocus cues has been analyzed by \cite{Kim2014} to select reliable pixels for depth estimation using a cost volume reconstructed from the light-field.
In \cite{Jeon2015}, a depth map estimation algorithm is presented using a multi-label optimization of a cost volume for lenslet-based light-field image.
They improved their depth estimation in \cite{Jeon2018c} by combining different matching costs and learned to automatically determine which combination performs better on the given input. 
Several aggregation costs have also been tested and evaluated in \cite{Williem2018}. %
Built upon the work of \cite{Tao2013}, the depth estimation is conducted in \citet{Wang2015} by treating occlusions explicitly in a photo-consistency model using the reconstructed central view.
To better account for correlation and dependencies within angular patches and spatial images, \cite{Zhang2019} proposed a two-step light-field depth estimation based on graph spectral analysis.\\
In another direction, recent work of \citet{Anisimov2019} aimed at reducing computational cost by leveraging a  semi-global matching strategy, instead of focusing on improving depth estimation.
Their method is based on pixel matching in \acp{SAI} with different similarity measurements for estimation of a dense depth map.\\
Another category of methods aims at leveraging the 4D light-field structure to extract features to be matched. 
Several descriptors have been proposed such as LiFF \citep{Dansereau2019} built upon SIFT, the binary descriptor introduced by \cite{Alain2020} built upon BOOM, or FDL-HSIFT \citep{Xiao2021} built upon Harris and SIFT in the scale-disparity space.\\

All the previous methods operate either on the light-field or on reconstructed \acp{SAI}. 
Those are easily available with a camera-array setup, sequential acquisition, or unfocused plenoptic cameras.
However, this is not the case for focused plenoptic cameras.
The latter setup leads to an ill-posed problem because depth information is required to reconstruct the light-field or the \acp{SAI} {\citep{Hog2017b,Liang2015,Wanner2011b}}.

\subsubsection{Based on epipolar plane images (EPIs)}

\ac{EPI} usually represents a 2D slice (in the spatial and angular dimensions) of the 4D light-field. 
From analysis of variations within this representation, we can infer depth information.
An overview and taxonomy of dense light-field depth estimation algorithms is available in \cite{Johannsen2017b}, mostly including methods working on reconstructed images or \acp{EPI}.

This structure was first analyzed by \cite{Bolles1987b} to retrieve depth by estimating lines in the \ac{EPI}, as the slope is inversely proportional to depth.
For focused plenoptic cameras, an algorithm to generate \ac{EPI} representation was proposed by \cite{Wanner2011b}.
They computed the full depth of field view for each lens type independently and then applied a merging algorithm to include the multi-focal aspect.
The latter representation was used by \cite{Wanner2012} who proposed a globally consistent framework using structure tensors to estimate the directions of feature pixels in the 2D \ac{EPI}.
In \cite{Yu2013}, geometric structures of 3D lines in ray space extracted from \ac{EPI} were explored. 
They encoded the line constraints to further improve the reconstruction quality.
An algorithm that computes dense depth estimation by combining both defocus and correspondence depth cues was introduced by \citet{Tao2013}. 
They latter included shading as a third clue to improve their depth estimation~\cite{Tao2015b}.
A method was proposed by \citet{Tosic2014b} to detect ray geometry in \acp{EPI} based on light-field scale and depth (Lisad) space transform. 
\cite{ShanXu2016d} implemented a three-step depth estimation using \acp{EPI}.
Latter methods are vulnerable to occlusions as they generate inconsistencies in the \acp{EPI}. 
\cite{Chen2014} explicitly tackled this issue by presenting a new light-field stereo matching algorithm that is capable of handling occlusion based on analysis of the angular statistics of the light-field.
More recently, \cite{Zhang2016b} introduced a \ac{SPO} to locate lines and calculate their orientations in an \ac{EPI} for local robust depth estimation.
According to the authors, \ac{SPO} has been demonstrated to be insensitive to occlusions, noise, spatial aliasing, or limited angular resolution. %
A new framework using \ac{SPO} was developed by \cite{Sheng2018} to locate lines in multi-orientation \acp{EPI}. \\

Similarly to \acp{SAI}-based methods, the \ac{EPI} representation can easily be retrieved for unfocused plenoptic cameras, but needs prior depth to be generated from focused plenoptic cameras. %
\subsubsection{Based on learning}

Deep learning methods have also been applied on light-field images, in particular in context of super-resolution.
The first deep convolutional neural network (CNN) that jointly optimizes angular and spatial super-resolution images from a pair of \acp{SAI} was proposed in \cite{Yoon2015}.
Generated \acp{SAI} were then used in a stereo matching-based depth estimation built-upon the method of \cite{Jeon2015}.
An end-to-end network was introduced by \cite{Ma2018a} using all \acp{SAI} allowing to capture both local and global features to generate a disparity map. 
To reduce the amount of input data, \cite{Shin2018a} developed a multi-stream fully CNN using only \acp{SAI} stacked in four angular directions to produce a disparity map.
\cite{Liu2020} presented a three-part neural network architecture that %
allows to take into account both parallax and ambiguity cues.
Recently, a new depth estimation based on unsupervised learning was proposed in \cite{Jin2021} to overcome the necessity of having depth maps as ground truth and to reduce the gap between simulated and real data.
Using \ac{EPI} representation, \cite{Johannsen2016} proposed a novel approach for depth estimation based on a learned dictionary which codes for disparity from \ac{EPI}. 
\cite{Heber2016} introduced a U-shaped fully CNN with skipped connections where inputs are \ac{EPI} representations and output is a disparity map.
Recently, \cite{Li2020b} proposed a pseudo-siamese neural network to estimate depth at each pixel, taking as input the vertical and the horizontal \ac{EPI} at this location.\\
Instead of explicitly including vision cues, \cite{Huang2018} proposed to model the light-field matching problem using an empirical Bayesian framework which better generalizes to different light-fields (dense, sparse, color, gray-scaled, etc.) to achieve better depth quality.\\

To the best of our knowledge, no learning-based methods are able to directly operate on raw plenoptic images, and therefore required \acp{SAI} or \acp{EPI} generation.

\subsubsection{Based on raw images}

To overcome the issues related to reconstruction of the \acp{SAI} or \acp{EPI}, several methods work directly with raw images. 
This is particularly suited for \textit{focused} plenoptic cameras (2.0), as each micro-image captures more spatial information than its unfocused counterpart.

With the arrival of commercial focused plenoptic cameras, \cite{Perwass2010c} proposed a methodology to estimate depth directly from raw images, later improved based on the camera model of \cite{Heinze2016b}.
Their method was based on triangulation from micro-images views employing a correlation technique just as in standard stereo-matching approaches.
But it required contrasted micro-images and sharp micro-images.
If the camera is calibrated and once each pixel has a depth estimate, sparse metric depth can be retrieved.
An automatic method is presented in \cite{Custodio2014} to estimate the depth of a scene based on multi-view geometry and ray back-tracing from detected salient points.
Use of points of interest matched between micro-images (SURF, SIFT and Harris) has also been investigated by \cite{Konz2016b} to estimate virtual depth by triangulation. 
However, the accuracy of the estimation was limited, and the number of features were too low to obtain a proper depth map. \\
In \cite{Noury2019}, metric depth estimation per micro-image was conducted.
Their method is inspired from \cite{Scharstein2002} but applied to micro-images from the raw plenoptic image.
It relies on a minimization process of the dense reprojection error of the reconstructed neighbors micro-images given a depth hypothesis following their projection model of the camera developed in \cite{Noury2017b}.\\
Depth estimation for the multi-focus plenoptic camera has been explicitly considered in \cite{Fleischmann2014}.
Their method, based on one depth estimation per micro-image, operates by regularizing a cost volume computed from a similarity measure between micro-images at different disparity hypotheses.
Disparities were then converted to virtual depths, according to the \ac{MLA} parameters.
To take into account the varying amount of defocus blur between micro-images of different types, they developed an adaptive strategy to select only certain candidates micro-images.
Similar to the previous work, \cite{Ferreira2016} used salient points detected with SIFT to select micro-images in which a search along epipolar line is performed. 
A specific lens selection scheme to improve robustness was proposed.
Finally, matched points were back-projected into virtual space to form a point cloud.
Virtual points are then reprojected and an average depth is attributed for every micro-image.
Lens selection strategies have also been addressed in \cite{Palmieri2017}.
Another depth estimation for the multi-focus plenoptic camera was proposed by \cite{Cunha2020}. 
Their method operates by first detecting edges in micro-images, which are second matched with neighbors micro-images. 
Finally, matched points were triangulated into virtual space, and then reprojected into metric space with calibration parameters available.
They addressed the issue of the different amount of blur during the matching by proposing a switching mechanism between intensity and LPQ domains.\\
\cite{Zeller2016i} introduced the first probabilistic depth estimation from raw plenoptic images obtained from a focused plenoptic camera. 
They addressed depth estimation as a multi-view stereo problem.
For each pixel having a sufficient gradient, virtual depth hypothesis is obtained by finding correspondences along epipolar lines in neighbors micro-images with local intensity error minimization.
Multiples hypotheses are merged in a Kalman-like fashion, allowing to associate a variance to the estimation.
To deal with the multi-focus aspect, they incorporated a term modeling the focus uncertainty. 

All previous solutions from the state-of-the-art using raw images considered blur as a drawback and designed specific strategies to select micro-images.
In contrast, leveraging the new camera model introduced by \cite{Labussiere2022calib}, we explicitly use the defocus information in our depth estimation process, taking into account both correspondence and defocus cues.
Micro-images are matched like in multi-views stereo matching problems to estimate disparities, which are then reconverted to metric depths thanks to the camera projection model explicitly including the defocus blur.%

\section{{Blur Aware} Plenoptic camera model}\label{sec:cameramodel}

\begin{figure*}[!t]
\centering
\includegraphics[height=5.7cm]{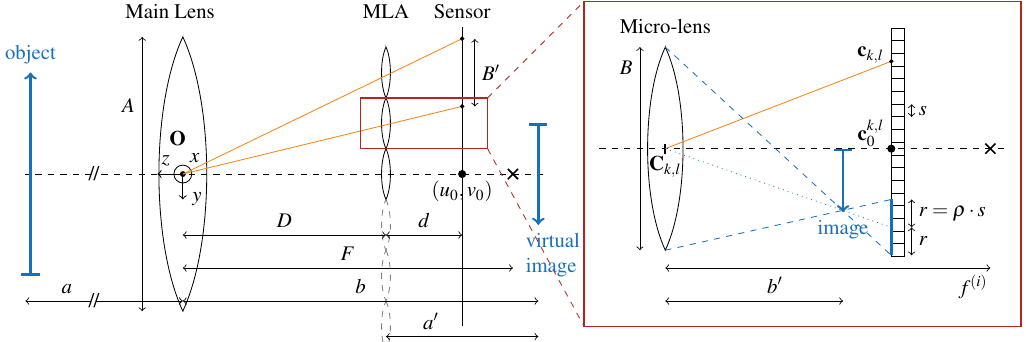}%
\caption{
	Blur Aware Plenoptic camera model \citep{Labussiere2022calib} with the notations used in this paper. Object points are projected by the main lens behind the \acf{MLA} into a virtual intermediate space, and then re-imaged by each micro-lens onto the sensor.%
}%
\label{fig:notations}\label{fig:model}%
\end{figure*}

We first briefly present the model of the focused plenoptic camera, especially the multi-focus case in \autoref{sec:projmodel}. %
The camera is composed of a main lens and photo-sensitive sensor with a \acf{MLA} in between, as illustrated in \autoref{fig:model}.
In the multi-focus configuration, the micro-lenses array consists of $\mltypenb$ different types of thin-lenses. 
Each micro-lens creates a \ac{MI} onto the sensor.
We consider the model presented by \cite{Labussiere2020blur} and extended in \cite{Labussiere2022calib}, explicitly including the micro-lenses focal lengths $\focal\mltype{i}$ and the defocus blur $\blurradiuspix$ in the direct projection model.
In order to relate image information to world information, we naturally derive here in \autoref{sec:invprojmodel} the inverse projection model.
{The inverse projection provides metric results up to a scale factor.
We will detail in \autoref{sec:depthcalibration} the proposed depth scaling mode and a calibration method to retrieve accurate depth.}

\subsection{Direct projection model}\label{sec:projmodel}

The blur aware plenoptic camera model links a scene point $\point\fworld = \bbm x&y&z&1\ebm\T$ to a \ac{BAP} feature $\feature_{\mlindexes} = \bbm u& v& \blurradiuspix & 1\ebm\T$ in homogeneous coordinates through each micro-lens $\left(\mlindexes\right)$ of type $(i)$.
The $\left(u,v\right)$ part encodes the point position in the image and the $\blurradiuspix$ variable encodes the radius of the defocus blur. More details of the defocus model will be given in \autoref{sec:blurmodel}.
The direct projection $\fun{\Pi_{\mlindexes}}$ from \cite{Labussiere2022calib} is then given by%
\begin{equation}\label{eq:completemodel}
\bbm u \\ v \\ \blurradiuspix \\ 1\ebm \propto
\bapprojm\left(i,\mlindexes\right)
\cdot 	\poseMLkl
\cdot 	\distor{
\Kthinlens\left(\Focal\right)
\cdot 	\poseLens
\cdot 	\point\fworld
}\tcomma%
\end{equation}
where
$\bapprojm\left(i,\mlindexes\right)$ is the blur aware plenoptic projection matrix through the micro-lens $\left(\mlindexes\right)$ of type $(i)$, and computed as%
\begin{equation}\label{eq:bapprojm}
\bapprojm\left(i,\mlindexes\right) = \mat{P}\!\left(\mlindexes\right)\cdot \Kthinlens\left(\focal\mltype{i}\right)\tdot
\end{equation}
$\mat{P}\!\left(\mlindexes\right)$ is a matrix that projects the 3D virtual point onto the sensor and takes into account the defocus blur.
$\Kthinlens\left(\focal\right)$ is the thin-lens projection matrix for the given {focal length}.
$\poseLens$ is the pose of the main lens with respect to the world frame 
and $\poseMLkl$ is the pose of the micro-lens $\left(\mlindexes\right)$ expressed in the camera frame.
The function $\funarg{\Distor}{\cdot}$ models the lateral distortion. 
The reader can refer to \cite{Labussiere2020blur,Labussiere2022calib} and supplementary materials for more details about the direct projection model.\\

\subsection{Inverse projection model}\label{sec:invprojmodel}

In order to relate image information to world information, we now inverse the projection model.
We back-project through a micro-lens $\left(\mlindexes\right)$ of type $(i)$ a \ac{BAP} feature $\point_{\mlindexes} = \bbm u & v & \rho & 1 \ebm\T$ into a point $\point\fworld = \bbm x & y & z &1 \ebm\T$ in object space.
The inverse projection $\fun{\Pi{}^{-1}_{\mlindexes}}$ is given by
\begin{equation}\label{eq:completeinvmodel}
\bbm x \\ y \\ z \\ 1\ebm \propto
\invposeLens
\cdot 	\invKthinlens\left(\Focal\right)
\cdot 	\invdistor{
\invposeMLkl
\cdot 	\invbapprojm\left(i,\mlindexes\right)
\cdot 	\point_{\mlindexes}
}\tdot
\end{equation}
\noindent\textbf{Blur radius computation.}
In practice, the blur radius $\blurradiuspix$ is linked to the virtual depth $\virtdepth$.
Virtual depth refers to relative depth value with respect to the \ac{MLA}. 
It is defined as the ratio between the signed object distance $\distobj'$ (as indicated on \autoref{fig:model}) and the sensor distance $\dist$, \ie
\begin{equation}\label{eq:virtdepth}
\virtdepth = -\frac{\distobj'}{\dist}\tdot
\end{equation}
The relation between $\blurradiuspix$ and $\virtdepth$ is given by
\begin{equation}\label{eq:blurvsvirtdepth}
\blurradiuspix =  m \cdot \virtdepth^{-1} + q_i\tcomma
\end{equation}
where $m$ and $q_i$ are known coefficients computed from the intrinsic parameters and depending on the micro-lens' type $(i)$ \citep[see Eq.~32]{Labussiere2022calib}.\\

In the end, we are now able to back-project a \ac{BAP} feature, \ie a pixel having a virtual depth into a point in object space. 
It allows us then to convert virtual depth to metric depth {up to a scale factor.}\\

\noindent\textbf{Inverse distortion model.}
Since the lateral distortion model of \cite{Labussiere2022calib} is not invertible, we add in this paper the inverse distortion $\invdistor{\cdot}$ explicitly in the camera model.
Inverse distortion is used to create an undistorted point $\point\fundistorted = \invdistor{\point\fdistorted} = \bbm x\fundistorted & y\fundistorted & z &1\ebm\T$ from a distorted point $\point\fdistorted = \bbm x\fdistorted & y\fdistorted & z &1\ebm\T$.
Note that distortions are applied in the 3D space to take into account the deviation from the thin-lens model, not in image space.
An efficient way to characterize the inverse distortion is to use a high order version of the Brown's model as shown in \cite{DeVilliers2008}. 
In particular, we use a model of the same order as the direct distortion model, and the mapping is expressed as
\begin{equation}\label{eq:invdistortion}
\left\{
\begin{aligned}
x\fundistorted =&~ x\fdistorted  \left(1 + Q_{\text{-}1} \varsigma^2 + Q_{\text{-}2} \varsigma^4 + Q_{\text{-}3} \varsigma^6 \right) &\text{[radial]}\\
&+ P_{\text{-}1} \left(\varsigma^2 + 2 x{}^2\fdistorted\right) + 2 P_{\text{-}2} x\fdistorted y\fdistorted& \text{[tangential]}\\
y\fundistorted =&~ y\fdistorted \left(1 + Q_{\text{-}1} \varsigma^2 +  Q_{\text{-}2} \varsigma^4 + Q_{\text{-}3} \varsigma^6\right) &\text{[radial]}\\
&+ P_{\text{-}2} \left(\varsigma^2 + 2 y{}^2\fdistorted\right) + 2 P_{\text{-}1} x\fdistorted y\fdistorted& \text{[tangential]}\\
\end{aligned}
\right.
\end{equation}
where $\varsigma^2 = {x\fdistorted^2+y\fdistorted^2}$.
The three coefficients for the radial component are given by 
$\set{Q_{\text{-}1}, Q_{\text{-}2}, Q_{\text{-}3}}$, 
and the two coefficients for the tangential by 
$\set{P_{\text{-}1}, P_{\text{-}2}}$.\\

\noindent\textbf{Calibration of the coefficients.} The optimization of the inverse coefficients is done only once, as a post-calibration step, such that
\begin{equation}
\underset{{\set{Q_{\text{-}1}, Q_{\text{-}2}, Q_{\text{-}3}, P_{\text{-}1}, P_{\text{-}2}}}}{\arg\min}~ \sum_{\point}\norm{\point - \invdistor{\distor{\point}}}^2\tcomma
\end{equation}
for a large number of samples $\point$ uniformly distributed in the virtual intermediate space (\ie after projection by the main lens).
The optimization is conducted with the Levenberg-Marquardt algorithm.
In practice, coefficients are initialized from the direct distortion coefficients, and less than fifteen iterations are sufficient to converge, with a root mean square error (RMSE) of approximately \SI{e-5}{\mm}.

\section{Depth scaling model and calibration}\label{sec:depthcalibration}\label{sec:scalemodel}

When mapping from the virtual space to the object space with the inverse projection model, the reconstructed objects are scaled up to a certain factor.
The scale factor grows approximately linearly as function of the distance with respect to the focus distance.
This phenomenon is present both on real and simulated data.
It is due to the limitations of the thin-lens model.
The proposed model describes efficiently the projective geometry for a point whose all rays inside the projection cone attain the retina, but this is not always the case.
Given specific configurations and with aperture corresponding to the $f$-number matching principle \citep{Perwass2010c}, not all rays from the cone of light reach the sensor, inducing a shift in the radiance. {See supplementary materials for more comments.}
Indeed, when simulating the imaging process with a large aperture, the retrieved depths from the simulated images do not suffer from a scale error.

Although not explicitly pointing out this issue, \citet{Zeller2016i} calibrated the mapping between virtual depth and metric depth in object space, by proposing three different models.
\citet{Heinze2016b} also noted the presence of a systematic error in the depth estimation process, even after correcting the offset induced by the thick-lens model.

In the following, we show, first, how we {empirically} model the depth scaling, second, how to measure and characterize the scaling error, and thus how to correct it in a post-calibration process.
Once the depth scaling has been calibrated, coordinates of the back-projected point (\ie from virtual to metric space) are corrected according to our proposed model, to achieve precise and accurate depth estimation.

\subsection{Scale correction model}

From experimental data {(both from several physical configurations and in simulation)}, we infer that the scaling error is function of the distance $z$, and that polynomial function is sufficient to fit the results.
We proposed then to model the scaling correction as a polynomial, noted $\funarg{\Gamma}{\cdot}$. 
The function $\fun{\Gamma}$ takes as input the $z$-component of the 3D point obtained by back-projection, and the point is then re-scaled by 
\begin{equation}\label{eq:scalefactor}
\gamma = \frac{\funarg{\Gamma}{z}}{z}\tcomma
\end{equation} 
\ie the corrected point ${\point}^*$ from a pixel $\point_{\mlindexes}$ having a virtual depth $\virtdepth$ is
\begin{equation}\label{eq:rescaledprojection}
{\point}^* = \gamma \cdot \bbm x \\y\\z\ebm\text{,~where~} \bbm x \\ y \\ z \\1 \ebm \propto \funarg{\Pi{}^{-1}_{\mlindexes}}{\bbm u\\v\\\blurradiuspix = m \cdot \virtdepth^{-1} + q_i\\1 \ebm}\tdot %
\end{equation} 
We found a quadratic model to be the best compromise between model simplicity and performances.
It allows to efficiently correct the scale error as shown in \autoref{sec:res}.

\subsection{Scaling error measurement}

To quantify the scale error, we use the relative \ac{MBE} to measure the relative difference of the distance between each pair of back-projected points $\norm{\bar{\point}_{i} - \bar{\point}_{j}}$ and the known distance $\norm{{\point}{}_{i} - {\point}{}_{j}}$ between these points. 
In practice, we use points corresponding to corners of a known-sized checkerboard.

First, we perform \ac{BAP} detection on raw image of a checkerboard.
Second, we back-project each \ac{BAP} feature $\feature{}_{\mlindexes}$ having a virtual depth $\virtdepth$ of the same cluster $\Frame{C}{}_i$, \ie corresponding to the same corner $\point_{i}$, using \autoref{eq:completeinvmodel}.
Third, we compute the centroid ${\bar{\point}{}_{i}}$ corresponding to the checkerboard corner $i$, as
\begin{equation}
{\bar{\point}{}_{i}} = \frac{1}{\# \Frame{C}{}_i}\cdot \sum_{\feature{}_{\mlindexes} \in \Frame{C}{}_i} \funarg{\Pi{}^{-1}_{\mlindexes}}{\feature{}_{\mlindexes}}\tcomma
\end{equation}
where $\#\Frame{C}_i$ is the number of observations in the cluster $\Frame{C}_i$.
Finally, the scale error $\fun{\varepsilon_{scale}}$, for a frame having $I \times J$ corners, is computed as 
\begin{equation}\label{eq:scaleerror}
\fun{\varepsilon_{scale}}
= \frac{1}{I\cdot J} \cdot \sum_{\left(i, j\right) \in I \times J}
\left(1 - 
\frac{
	\norm{\bar{\point}_{i} - \bar{\point}_{j}}
}
{
	\norm{{\point}{}_{i} - {\point}{}_{j}}
}
\right)\tdot
\end{equation}

\subsubsection{Depth scaling calibration}

At this point, the camera intrinsic parameters, the relative blur coefficient and the inverse distortion coefficients have been calibrated. %
We propose then a post-calibration process for the scaling correction based on non-linear optimization of the scale error over several checkerboard raw plenoptic images.
Let $\Xi = \set{\gamma_0, \gamma_1, \gamma_2}$ be the set of parameters to optimize, such that $\funarg{\Gamma}{z} = \gamma_2 z^2 + \gamma_1 z + \gamma_0$.
The cost function $\Theta\!\left(\Xi\right)$ is expressed as the sum over each frame $n$ of the scale errors $\fun{\varepsilon_{scale}}$, \ie
\begin{equation}
\Theta\!\left(\Xi\right) = 
\frac{1}{I J N} \cdot
\sum_{n} \sum_{\left(i, j\right) \in I \times J}
\left(1 - 
\frac{
	\norm{\gamma_i\cdot\bar{\point}{}^n_{i} - \gamma_j\cdot\bar{\point}{}^n_{j}}
}
{
	\norm{{\point}{}^n_{i} - {\point}{}^n_{j}}
}
\right)\tcomma
\end{equation}
where $N$ is the number of frames and $I \cdot J$ is the number of checkerboard corners.
Each point $\bar{\point}{}^n_{i}$ is re-scaled by $\gamma_i$ as defined in \autoref{eq:scalefactor} and \autoref{eq:rescaledprojection}.
The optimization is conducted using the Levenberg-Marquardt algorithm.

\section{Linking disparity and defocus blur}\label{sec:linkdefocusdisp}

In order to obtain metric depth, we first estimate the virtual depth 
from disparity, and defocus is a complementary cue to resolve the inherent ambiguity of stereo matching as shown in \cite{Held2012,Schechner2000}.
The goal is then to improve disparity estimation for defocus stereo images via compensating the mismatch of focus thus integrating both correspondence and defocus cues.

\subsection{Defocus and relative blur models}\label{sec:blurmodel}

{We first recall how blur and relative blur between micro-images are modeled as presented in \cite{Labussiere2022calib}.}\\

\noindent\textbf{Defocus blur models.}
We modeled the geometric blur using the \acf{CoC}. The blur radius $\rho$ and its metric counter part $\blurradiusmm$ as in \cite[Eq.~6]{Labussiere2022calib}.
From a signal processing point of view, the response of an imaging system to an object out-of-focus can be modeled by the \acf{PSF}. 
Let $\img{x,y}$ be the observed blurred image of an object at a constant distance.
The image is the result of the convolution of the \ac{PSF}, noted $\Psf$, with the in-focus image, $\Img^*\!\left(x,y\right)$, as define in \cite[Eq.~7]{Labussiere2022calib}.
The spread parameter $\sigma$ of the \ac{PSF} is proportional to the blur circle radius $\blurradiuspix$.
Therefore, we can write
\begin{equation}\label{eq:sigma2rho}
\sigma \propto \blurradiuspix \Leftrightarrow \sigma = \kappa \cdot \blurradiuspix
\end{equation}
where $\kappa$ is a camera constant that is determined by calibration, using the method presented in \cite{Labussiere2022calib}.
Note that the spatially-variant spread parameter $\sigma$ thus depends on the object distance, \ie blur and depth are linked. \\

\noindent\textbf{Micro-lenses relative blur.}
A point imaged by two different micro-lenses of type $(i)$ and $(j)$ will have different blur radii, \ie the resulting images will have different spread parameters for the \ac{PSF} model, such that
\begin{equation}\label{eq:imgformation}
\left\{
\begin{aligned}
~\Img_{(i)}\!\left(x,y\right) &= \convp{\psf_{(i)}}{\Img^*\!\left(x,y\right)} + \fun{n}_{(i)}\!\left(x,y\right)\\
~\Img_{(j)}\!\left(x,y\right) &= \convp{\psf_{(j)}}{\Img^*\!\left(x,y\right)} + \fun{n}_{(j)}\!\left(x,y\right)\tcomma
\end{aligned}
\right.
\end{equation}
where $\Img^*\!\left(x,y\right)$ is the latent in-focus image, {and $\fun{n}\!\left(x,y\right)$ is the image noise}\footnote{%
{
		Noise can be omitted under the assumption that, $\convp{\fun{n}_{(i)}\!\left(x,y\right)}{\psf_{(j)}} \approx \convp{\fun{n}_{(j)}\!\left(x,y\right)}{\psf_{(i)}} \approx 0$.
		This assumption does not hold when exactly one of the images is in focus. However if we blur both images by a small amount then we can expect the assumption to be valid. The approximation error is small when both images have large blur.%
	}
}.
We then use the equally-defocused representation \citep[Eq.~36]{Labussiere2022calib} by applying additional blur to the relatively in-focus micro-image,
\begin{equation}\label{eq:relblurimgformation}
\left\{
\begin{aligned}
~\Img_{(i)}\!\left(x,y\right) &\simeq \convp{\psf_{r}}{\Img_{(j)}\!\left(x,y\right)} &\text{if } \sigma_{(i)} \ge \sigma_{(j)} \\
~\convp{\psf_{r}}{\Img_{(i)}\!\left(x,y\right) } &\simeq \Img_{(j)}\!\left(x,y\right) &\text{if } \sigma_{(i)} < \sigma_{(j)}
\end{aligned}
\right.\tdot
\end{equation}
$\psf_{r}$ is the relative blur kernel, of spread parameter $\sigma_r$, applied to either one of the views such that both are equally-defocused.

\subsection{Link between disparity and the relative blur}

As highlighted in \cite{Chen2015}, while estimating depth from a defocus stereo configuration, both spatially-variant blur and disparity provide the inference for depth information.
Establishing the visual correspondence across two images must take both disparity and blur into account. \\

\noindent\textbf{Defocus stereo images configuration.}
In standard stereo matching, two focused rectified stereo images are used to determine disparity. 
The left image $\Img_{(i)}$ and right image $\Img_{(j)}$ are related with the spatially-variant disparity $\vect{\disparity} \in \Reals^2$, and thus the correspondence between the two images can be modeled by
\begin{equation}\label{eq:disprelation}
\Img_{(i)}\!\left(\point\right) = \Img_{(j)}\!\left(\point+\vect{\disparity}\right)
\end{equation}
where $\point=\left(x,y\right)$ is the spatial index of a pixel.
Taking into account blur, we model each image as the convolution of the in-focus image with a blur kernel %
as in \autoref{eq:imgformation} and we consider the equally defocused images as in \autoref{eq:relblurimgformation}.
Therefore, injecting the disparity in the equally defocused model, the correspondence is given by
\begin{equation}\label{eq:blurdisparity}
\left\{
\begin{aligned}
~\Img_{(i)}\!\left(\point\right) &\simeq \convp{\psf_{r}}{\Img_{(j)}\!\left(\point+\vect{\disparity}\right)} &\text{if } \sigma_{(i)}\!\left(\point\right) \geq \sigma_{(j)}\!\left(\point+\vect{\disparity}\right) \\
~\convp{\psf_{r}}{\Img_{(i)}\!\left(\point\right) } &\simeq \Img_{(j)}\!\left(\point+\vect{\disparity}\right) &\text{if } \sigma_{(i)}\!\left(\point\right) < \sigma_{(j)}\!\left(\point+\vect{\disparity}\right)
\end{aligned}
\right.\tdot
\end{equation}

\noindent\textbf{Relative blur as function of the virtual depth.}
From the blur radius formula, the relative blur can be approximated by a linear function of the disparity (\ie of the inverse virtual depth up to a factor), such that 
\begin{equation}\label{eq:relblurvsvirtdepth}
\Delta r^2 
= {r_{(i)}^2 - r_{(j)}^2} 
\approx m_{i,j} \cdot \virtdepth^{-1} + q_{i,j}
\text{~~~~[\si{\mm\squared}]}
\end{equation}
with
\begin{equation}\label{eq:mij}
m_{i,j} = \frac{B^2d}{2}\cdot\left(\frac{1}{{a_0}^{(i)}} - \frac{1}{{a_0}^{(j)}}\right) 
\end{equation}
and
\begin{equation}\label{eq:qij}
q_{i,j} = \frac{B^2d^2}{4}\cdot\left(\left(\frac{1}{{a_0}^{(i)}}\right)^2 - \left(\frac{1}{{a_0}^{(j)}}\right)^2 \right)\tcomma
\end{equation}
where $B$ is the micro-lens diameter and ${a_0}^{(i)}$ is the distance to the plane of focus of the type $(i)$ micro-lens, computed as
\begin{equation}
{a_0}^{(i)} = \frac{\distmlasensor\focal\mltype{i}}{\distmlasensor - \focal\mltype{i}}\tdot
\end{equation}
All the parameters are known thanks to the camera calibration.
Finally, the spread parameter $\sigma_r$ is computed using \autoref{eq:relblurvsvirtdepth} and \autoref{eq:sigma2rho}, such that
\begin{equation}\label{eq:pixelblurvsvirtdepth}
\sigma_r = \kappa \cdot \frac{1}{s} \cdot {\abs{\Delta r^2}}^{\frac{1}{2}}\tdot
\end{equation}
Note that the approximation of \autoref{eq:relblurvsvirtdepth} is exact when considering that micro-lenses are parallel to the sensor plane.
When dealing with micro-lenses in the same local neighborhood, the $z$-shift inducing a slight difference between the virtual depths can be neglected, and with orthogonal approximation, the relation stands.

\section{Blur Aware Depth Estimation (BLADE)}\label{sec:blade}

We make the hypothesis that the camera is calibrated and we have access to the intrinsic parameters.
We propose a process based on area matching techniques to estimate a raw depth map $\mathcal{D}$ directly from raw plenoptic images.
Two variations are considered: 1) \textit{coarse} estimation, \ie one depth per micro-image; and 2) \textit{refined} estimation, \ie one depth per pixel.
\autoref{alg:blade} summarizes the estimation process in the refined case.
A new residual error is formulated to leverage blur information for depth estimation using a multi-focus plenoptic camera. 
The computation is illustrated in \autoref{fig:blade}.
Example of depth maps obtained by our method are illustrated in \autoref{fig:coarsedm}.
Let an observation be a pair of micro-images such that the \emph{reference} {$\Img_{(i)}$} is the most defocused and the \emph{target} {$\Img_{(j)}$} is the micro-image to be equally defocused. \\

\noindent\textbf{Disparity.} 
The disparity is given by $\disparity = \norm{\vect{\disparity}}$ where 
$\vect{\disparity} \in \Reals^2$ is obtained at a virtual depth hypothesis $\virtdepth$ usually using the following relation
\begin{equation}\label{eq:falsedisp}
\vect{\disparity} = \frac{1}{\virtdepth}\cdot \vect{\baseline} \text{~~~with~~~} \vect{\baseline} = \left(\mlcenter^* - \mlcenter\right)\tcomma
\end{equation}
where $\mlcenter^*, \mlcenter$ are respectively the centers of the reference and target micro-lenses in the \ac{MLA} plane, and $\vect{\baseline}$ is the baseline. 
This relation gives the disparity in case of orthogonal projection of micro-lens center to micro-image center. 
To take into account the deviation of the micro-image centers, the corrected disparity $\disparity'$ in micro-image space is given by
\begin{equation}\label{eq:truedisp}
\vect{\disparity'} = \frac{\left(1-\lambda\right)\cdot \virtdepth + \lambda}{\virtdepth} \cdot \vect{\baseline'}
\end{equation}
with
\begin{equation}
\vect{\baseline'} = \lambda \cdot\vect{\baseline} = \left(\micenter^* - \micenter\right)\tcomma
\end{equation}
where $\lambda = \distlensmla / \left(\distlensmla+\distmlasensor\right)$, 
and $\micenter^*, \micenter$ are respectively the centers of the reference and target micro-images defining the baseline $\vect{\baseline'}$ in image space. \\ %

\begin{figure}[!t]%
\caption{Process of computing a refined depth map $\mathcal{D}\!\left(x,y\right)$ using our Blur Aware Depth Estimation (BLADE) framework.}\label{alg:blade}
\renewcommand\algorithmicrequire{\textbf{Input:}}
\renewcommand\algorithmicensure{\textbf{Output:}}
\hrule
\begin{algorithmic}[1]
	\Require Raw image, Camera model
	\Ensure Refined depth map $\mathcal{D}\!\left(x,y\right)$
	\ForAll {micro-image $\Img$}
	\State retrieve default neighborhood $\neighb{\Img}$
	\ForAll {pixel $\left(x, y\right)$ with enough texture \textbf{in} $\Img$}
	\State compute initial virtual depth $\virtdepth_0$ \Comment{\autoref{eq:depthestimateefine}}
	\State $\mathcal{D}\!\left(x,y\right) \gets \virtdepth_0$
	\EndFor
	
	\State update neighborhood $\neighb{\Img, \virtdepth_0}$
	\ForAll {pixel $\left(x, y\right)$ with enough texture \textbf{in} $\Img$}
	\State compute virtual depth $\hat{\virtdepth}$ \Comment{\autoref{eq:depthestimateefine}}
	\State $\mathcal{D}\!\left(x,y\right) \gets \hat{\virtdepth}$
	\EndFor			
	\EndFor
	\State convert virtual to metric using $\Proj{}^{-1}_{\mlindexes}$	\Comment{\autoref{eq:completeinvmodel}}
\end{algorithmic}%
\hrule
\end{figure}%

\noindent\textbf{Matching problem.}
Before triangulation, we need to know which pixel in neighbors micro-images are images of the same object point. 
The correspondence is modeled by an affine warping function $\funarg{\omega}{\Img, \vect{\disparity}}$ at the disparity hypothesis $\vect{\disparity} \in \Reals^2$ along the epipolar line that is applied to rectify the image $\Img$, such that
\begin{equation}\label{eq:warp}
\funarg{\omega}{\Img, \vect{\disparity}}\left(\point\right) = \Img\!\left(\point+\vect{\disparity}\right)\tdot
\end{equation}
Pixel intensities are interpolated using bilinear-interpolation. 
Pixels which are not reprojected are set to \num{0}.
We can compare the warped target image with the reference image to estimate if the disparity hypothesis is correct.
This is done by calculating the sum of absolute differences (SAD). \\

\begin{figure*}[!t]
\centering
\includegraphics[height=6.3cm]{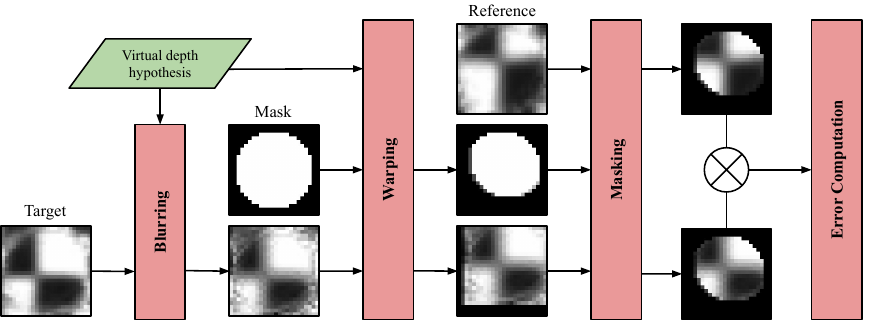}%
\caption{
	Process of computing the similarity residual error in our BLADE framework. 
	First, the target is equally-defocused given the virtual depth hypothesis.
	Second, the mask and the target are warped at the corresponding disparity.
	Third, the reference and the target are masked and compared to compute the similarity error.
}\label{fig:blade}
\end{figure*}

\noindent\textbf{Mask.}
To deal with circular micro-image of center $\micenter$ and radius $\varrho$, we define a mask image $\mathcal{M}$ by
\begin{equation}\label{eq:mask}
\mathcal{M}\!\left(\point\right) =
\left\{
\begin{aligned}
0 &&\text{if } \norm{\point - \micenter} > \varrho - b \\
1 &&\text{if } \norm{\point - \micenter} \leq \varrho - b
\end{aligned}
\right.
\end{equation}
where $b$ is the margin border of the micro-image. In our experiments, we used $b=1.5~\si{pixel}$ to minimize the vignetting effect. %
The final mask $\mathcal{M}^*$ is given as the intersection of the circular mask and the warped circular mask (see \autoref{fig:blade}), representing the common pixels between the reference and the target at the given disparity hypothesis, i.e.,
\begin{equation}
\mathcal{M}^* = \mathcal{M} \circ \funarg{\omega}{\mathcal{M}, \vect{\disparity}}\tcomma
\end{equation}
where $\circ$ is the element-wise matrix multiplication. \\ %

\noindent\textbf{Blur equalization.}
One specificity of the multi-focus plenoptic camera is that for a same portion of a scene observed in micro-images using different focal lengths, these micro-images will demonstrate different amounts of blur.
To compensate for the blur mismatch between the reference and the target, we use the equally defocused representation by adding supplemental blur to the target image.
The spread parameter $\sigma_r$ is obtained from \autoref{eq:pixelblurvsvirtdepth}.
To avoid dealing with micro-image borders while adding the relative blur, we use the S-Transform formulation from \cite{Subbarao1994} to compute the convolution with the blur kernel.
The equally defocused target image $\bar{\Img}$ is then computed as 
\begin{equation}
\bar{\Img} = \convp{\psf_{r}}{\Img} {~\approx~} \Img + \frac{{\sigma_r}^2}{4} \cdot \nabla^2\Img\tcomma
\end{equation}
where $\sigma_r$ is the spread parameter of the blur kernel, and $\nabla^2$ is the Laplacian operator. \\

\noindent\textbf{Similarity error computation.}
The similarity residual error is computed as the normalized SAD between the masked reference image and the masked equally-defocused matched target image.
It is expressed as 
\begin{align}\label{eq:simerror}
\funarg{\varepsilon_{sim}}{\Img_{(i)}, \Img_{(j)}, \vect{\disparity}} 
&= \eta \cdot \sum_{\point} 
\abs{ \Img_{(i)}\!\left(\point\right) - 
\funarg{\omega}{\bar{\Img}_{(j)}, \vect{\disparity}}\!\left(\point\right)  }
\cdot \mathcal{M}^*\!\left(\point\right) \nonumber \\
&= \eta \cdot \norm{\left(\Img_{(i)} - 
\funarg{\omega}{\bar{\Img}_{(j)}, \vect{\disparity}}\right) \circ \mathcal{M}^*}_1 \tcomma
\end{align}
with $\eta$ being the normalization factor defined as the sum of the common pixels between the target and the reference.
This factor is given by
\begin{equation}\label{eq:normerror}
\eta^{-1} = \sum_{\point}\mathcal{M}^*\!\left(\point\right) = \norm{\mathcal{M}^*}_1 \tcomma
\end{equation}
where $\norm{\cdot}_1$ is the entry-wise matrix $\ell_1$-norm, \ie $\norm{\mat{A}}_1 = \sum_{{i,j}}\abs{a_{i,j}}$.
Normalization of the error is required as the number of pixels to take into account varies with the disparity, and therefore with the virtual depth and the pair of micro-images.
For a same virtual depth hypothesis, according to the pair of micro-images, the disparity will change. \\

\noindent\textbf{Cost computation.}
For a micro-image $\Img_{(i)}$, the cost is the weighted sum of all the errors computed for each micro-image $\Img_{(j)}$ in its neighborhood $\neighb{\Img_{(i)}, \virtdepth}$ at the given virtual depth hypothesis $\virtdepth$, i.e., the cost $\Theta\!\left(\Img_{(i)}, \virtdepth\right)$ is 
\begin{equation}\label{eq:cost}
\frac{1}{W} 
\cdot \sum_{\Img_{(j)} \in \neighb{\Img_{(i)}, \virtdepth}} 
\funarg{w}{\Img_{(i)}, \Img_{(j)}} \cdot \funarg{\varepsilon_{sim}}{\Img_{(i)}, \Img_{(j)}, \vect{\disparity}}\tcomma
\end{equation}
with $\funarg{w}{\Img_{(i)}, \Img_{(j)}}$ being a weight function, and $W$ being the total weight computed as
\begin{equation}\label{eq:totalweight}
W = \sum_{\Img_{(j)} \in \neighb{\Img_{(i)}, \virtdepth}} \funarg{w}{\Img_{(i)}, \Img_{(j)}}\tdot
\end{equation}
In our experiment, we define the weight as constant. \\

\noindent\textbf{Initialization.} The number of \acp{MI} that see the same scene points in the neighborhood $\neighb{\Img, \virtdepth}$ of the considered \ac{MI} depends on the virtual depth hypothesis. 
So, we first coarsely initialize $\virtdepth_0$ from \acp{MI} of the same type (here, at baseline $\baseline = 2 \cdot \sin\frac{\pi}{3}$ in case of a hexagonal \ac{MLA} organization with three types of micro-lenses).
We use then $\virtdepth_0$ to retrieve the correct neighborhood, and we restrict the search of the optimal value to $\virtdepth_0 \pm N$, with $N = 1.96$ in our experiments.\\

\noindent\textbf{Coarse depth estimation.}
Under the hypothesis of one depth per micro-image, corresponding to locally planar approximation, the coarse depth map $\mathcal{D}\!\left(k, l\right)$ is estimated as follows. For each micro-image $\Img$, virtual depth estimation is conducted by a minimization of the latter cost function in an optimization process, such that
\begin{equation}\label{eq:depthestimatecoarse}
\hat{\virtdepth} = \arg\min_{\virtdepth} \Theta\!\left(\Img, \virtdepth\right)\tdot
\end{equation}
As the function is in 1-D, we use the Golden Search Section (GSS) algorithm \citep{Kiefer1953} to find the minimum with the desired precision.
To improve time computation, only \acp{MI} with sufficient amount of texture are considered, such that
\begin{equation}
\funarg{std}{\Img, \mathcal{M}} > t_{\mathrm{c}}\tcomma
\end{equation}
where $\funarg{std}{\cdot, \cdot}$ is the standard deviation of the pixels intensity $\point \in \Img \mid \mathcal{M}\!\left(\point\right) \neq 0$, and $t_{\mathrm{c}}$ is a threshold to reject non-textured area.
In our experiments, we set $t_{\mathrm{c}} = 5$.
Example of coarse virtual depth map is given in \autoref{fig:coarsedm}(a).\\

\noindent\textbf{Refined depth estimation.}
Under the hypothesis of one depth per pixel, a refined depth map $\mathcal{D}\!\left(x, y\right)$ is computed. 
The virtual depth estimation is conducted in a similar fashion as for the coarse estimation.
For a pixel $\point = \left(x, y\right)$, errors and costs are computed the same way as previously but considering only the result within a window $\mathcal{W}$ extracted around $\point$.
In our experiments, we use a window of size $5 \times 5~\si{pixel}$.
The similarity residual error $\funarg{\varepsilon_{sim}}{\Img_{(i)}, \Img_{(j)}, \vect{\disparity}, \point}$ is then given by 
\begin{equation}\label{eq:simerrorpixel}
\norm{
\mathcal{W}\!\left( \point,\mathcal{M}^*\right)
}_1^{-1}  
\cdot 
\norm{
\mathcal{W}\!\left( \point,
\left(\Img_{(i)} - \funarg{\omega}{\bar{\Img}_{(j)}, \vect{\disparity}}\right) \circ \mathcal{M}^*
\right)
}_1 \tdot
\end{equation}
The cost at a pixel $\point$ having a sufficient contrast, \ie such that it  verifies
\begin{equation}
\funarg{std}{\mathcal{W}\!\left(\point, \Img\right), \mathcal{W}\!\left(\point, \mathcal{M}\right)} > t_{\mathrm{c}}\tcomma
\end{equation}
is given by $\Theta\!\left(\Img_{(i)}, \virtdepth, \point\right)$ as
\begin{equation}\label{eq:costpixel}
\frac{1}{W} 
\cdot \sum_{\Img_{(j)} \in \neighb{\Img_{(i)}, \virtdepth}} 
\funarg{w}{\Img_{(i)}, \Img_{(j)}} \cdot \funarg{\varepsilon_{sim}}{\Img_{(i)}, \Img_{(j)}, \vect{\disparity}, \point}\tcomma
\end{equation}
where the weighs are defined as in \autoref{eq:totalweight}.
Finally, we compute the virtual depth $\hat{\virtdepth}$ at each pixel $\point$ such that
\begin{equation}\label{eq:depthestimateefine}
\hat{\virtdepth} = \arg\min_{\virtdepth} \Theta\!\left(\Img, \virtdepth, \point\right)\tdot
\end{equation}
Example of refined virtual depth map is given in \autoref{fig:refineddm}(b). \\

\begin{figure}[!t]
\centering
\includegraphics[width=\linewidth]{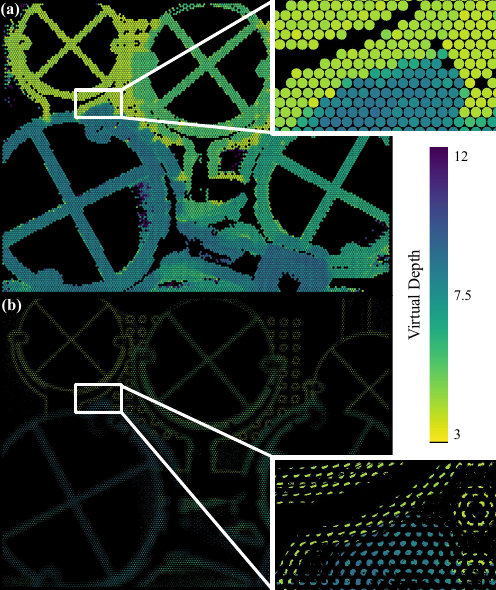}
\caption{Examples of raw virtual depth maps obtained by our BLADE framework with a zoom on an occluded area.
	\textbf{(a)} Coarse virtual depth map $\mathcal{D}\!\left(k,l\right)$, with one estimation per micro-image.
	\textbf{(b)} Refined virtual depth map $\mathcal{D}\!\left(x,y\right)$, with one estimation per pixel. More details can be captured per micro-image, but the map is sparser.\vspace*{-2em}
}\label{fig:coarsedm}\label{fig:refineddm}
\end{figure}

\noindent\textbf{Converting virtual to metric depth.}
At the end of the depth estimation process, we have a virtual depth estimate associated to each pixel.
To obtain a metric depth, we use the inverse projection model given in \autoref{eq:completeinvmodel} taking into account depth scaling as in \autoref{eq:rescaledprojection}, where $\blurradiuspix$ is computed from the virtual depth by \autoref{eq:blurvsvirtdepth}. %
\section{Experimental setup}\label{sec:setup}

\begin{figure}[!t]
\centering
\includegraphics[]{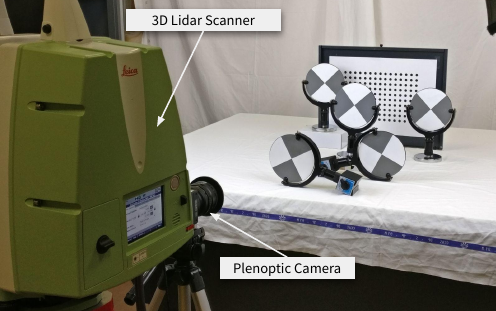}
\caption{
	Our \texttt{Raytrix R12} multi-focus plenoptic camera in our experimental setup, capturing a scene for 3D reconstruction. A \texttt{Leica ScanStation P20} allows us to capture a colored point cloud that can be used as ground truth data.\vspace*{-1em}
}\label{fig:setup}
\end{figure}

To validate our blur aware depth estimation framework, we proceeded as follows:
first, we analyzed the %
scaling error modelings;
second, we compared our method on relative depth estimation with state-of-the-art methods, including the \texttt{Raytrix} software, corresponding to the model of \citet{Heinze2016b}, and estimation with the model of \citet{Noury2017b} using only the disparity;
finally, we evaluated the depth estimation on real-world 3D complex scenes with ground truths acquired with a lidar.
Our experimental setup is illustrated in \autoref{fig:setup}.

\subsection{Hardware environment}

\noindent\textbf{Camera setup.} For our experiments we used a \texttt{Raytrix R12} color 3D-light-field-camera, with a \ac{MLA} of F/2.4 {aperture}.
The camera is in Galilean configuration, \ie the micro-lens focal lengths are greater than the distance \ac{MLA}-sensor.
The mounted lens is a \texttt{Nikon AF Nikkor F/1.8D} with a \SI{50}{\milli\meter} {focal length}.
The \ac{MLA} organization is hexagonal row-aligned, and composed of $176\times152$ (width $\times$ height) micro-lenses with $\mltypenb=3$ different types.
The sensor is a \texttt{Basler beA4000-62KC} with a pixel size of $\pixelsize= 0.0055$ \si{\milli\meter}.
The raw image resolution is $4080\times3068~\si{pixel}$.
We used four focus distance configurations, with $\focusdist \in \set{450, 1000, 2133, \infty}$ \si{\mm}.
Note that when changing the focus setting, the main lens moves with respect to the block \ac{MLA}-sensor.
{Therefore, we can consider that is equivalent to four physical different cameras, allowing us to validate our methodology on multiple setups.}\\

\noindent\textbf{Relative depth setup.} The camera is mounted on a linear motion table with micro-metric precision {as presented in \cite{Labussiere2020blur}}.
The target plane is orthogonal to the translation axis, and the camera optical axis is aligned with this axis.
Images with known relative translation between each frame are then used to estimate depths and compared to the ground truth, {for quantitative evaluation.} \\

\noindent\textbf{3D scenes setup.} We used a 3D lidar scanner, a \texttt{Leica ScanStation P20 (LP20)}, that allowed us to capture a color point cloud with high precision that can be used as {metric} ground truth {3D} data. 
The \texttt{LP20} is configured with no HDR and with a resolution of \SI{1.6}{\mm} at \SI{10}{\m}.

\subsection{Software environment}

All images have been acquired using the \texttt{MultiCamStudio} free software (v6.15.1.3573) of the \texttt{Euresys} company.
We set the shutter speed to $5$ \si{ms}.
For \texttt{Raytrix} data, we use their proprietary software \texttt{RxLive} (v4.0.50.2) to calibrate the camera, and compute the depth maps used in the evaluation.
Our source code has been made publicly available at
\url{https://github.com/comsee-research/libpleno}, and \url{https://github.com/comsee-research/blade}.

\subsection{Datasets}

\noindent\textbf{R12-A, B, C.} For the relative depth evaluation, we used the datasets presented in \cite{Labussiere2020blur}.
We evaluated then our depth estimation framework for focus distances $\focusdist \in \set{450, 1000, \infty}$ \si{\mm} (i.e., datasets \texttt{R12-A,B,C} respectively), {representing then three different working depth ranges.}

\noindent\textbf{R12-E, ES, ELP20.} For the 3D scenes evaluation, we introduced a new dataset, namely \texttt{R12-E}, corresponding to a {new} focus distance $\focusdist = 2133~\si{\mm}$.
The camera has been calibrated using {\texttt{Compote}}~\citep{Labussiere2022calib}.
From this configuration, we created two sub-datasets:
1) a simulated dataset built upon our own simulator based on raytracing\footnote{%
	Available at \url{https://github.com/comsee-research/prism}
} to generate images with known absolute position, named \texttt{R12-ES};
2) a dataset composed of several {real-world} 3D scenes with {metric} ground truth acquired with the \texttt{LP20}, for object distances ranging from \SIrange{400}{1500}{\mm}, {exploiting then the whole distance range offered by the camera setup}. %
The latter dataset, named \texttt{R12-ELP20}, includes fives scenes: 1) a scene for extrinsics calibration, containing checker corner targets, named \texttt{Calib}; 2) two scenes containing textured planar objects, named \texttt{Plane-1} and \texttt{Plane-2}; 3) and two scenes containing various figurines, named \texttt{Figurines-1} and \texttt{Figurines-2}.
Each scene is composed of: a colored point cloud (with spatial $(x,y,z)$ information, color information $(r,g,b)$, and intensity information) in format \texttt{.ptx}, \texttt{.pts} and \texttt{.xyz}; 3D positions of the targets in the lidar reference frame; two raw plenoptic images in rgb color and two raw plenoptic images in bayer; finally, photos and labels of the scene.

Our datasets have been made publicly available, and can be downloaded from our github repository with their descriptions. %
{We believe it is a first step toward helping future research to evaluate their method on metric space.}

\subsection{Lidar-camera calibration}

In order to transform the lidar point cloud data $\mat{P}_{{l}}$ in the same reference frame as the camera, we calibrate the extrinsic parameters between those frames, \ie the transformation ${}^{c}\Transform_{{l}} \in \SE{3}$ such that $\mat{P}\fcam = {}^{c}\Transform_{{l}}\mat{P}_{{l}}$, where $\mat{P}\fcam$ is the point cloud data expressed in the camera frame. 
The calibration is a four-steps process.\\
First, using the \texttt{Leica} station, we acquired a point cloud of a scene containing calibration targets, and associated the 3D coordinates $\point_{{l}}$ of the corners manually from the point cloud. 
This set of corners forms a points constellation, noted $\mathcal{C}_{{l}}$.
Second, a raw plenoptic image of the same scene is acquired with the plenoptic camera, and \ac{BAP} features $\lbrace\point_{\mlindexes}\rbrace$ are extracted. 
The features are clustered and associated to the points constellation.
For each cluster of observations, the barycenter is computed.
Those barycenters are initial estimates of the projections of the points constellation through the main lens using a standard pinhole model.
Third, the initial transformation ${}^{c}\hat{\Transform}_{{l}}$ is estimated using the \ac{PnP} algorithm \cite{Kneip2011b}, like in classic pinhole imaging system.
Finally, the transformation ${}^{c}\Transform_{{l}}$ is refined by minimizing the reprojection error of the points constellation, such that
\begin{equation}
\underset{{}^{c}\Transform_{{l}}}{\arg\min}~ \sum_{\point_{{l}} \in \mathcal{C}_{{l}}} \sum_ {k,l} \norm{\point_{\mlindexes} - {\Proj_{\mlindexes}\left({}^{c}\Transform_{{l}}\point_{{l}}\right)}}^2 \tdot
\end{equation}
The optimization is conducted using the Levenberg-Marquardt algorithm.
The point cloud data $\mat{P}_{{l}}$ can now be expressed in the camera frame.
It is thus used as ground truth for quantitative evaluations.

\section{Results and discussions}\label{sec:res}

In the following, a relative error $\varepsilon_z$ for a known displacement $\delta_z$ is computed as the mean absolute relative difference between the estimated displacement $\hat{\delta}_z$ and the ground truth, for each pair of frames $\left(n, m\right)$ separated by a distance $\delta_z$, \ie
\begin{equation}
\mathrm{\varepsilon}_z\!\left(\delta_z\right) = \eta^{-1} \sum_{\left(n,m\right) \mid z_m - z_n = \delta_z} \frac{|{\delta_z - \hat{\delta_z}}|}{\delta_z}\tcomma
\end{equation}
where $\hat{\delta}_z = \hat{z}_m - \hat{z}_n$,
and $\eta$ is a normalization constant corresponding to the number of frames pairs. 
Except if indicated otherwise, $\hat{z}$ is the median of the depth estimates of the considered frame. %

\subsection{Depth scaling error analysis}

\begin{figure}[!t]
	\centering
	\includegraphics[width=\linewidth]{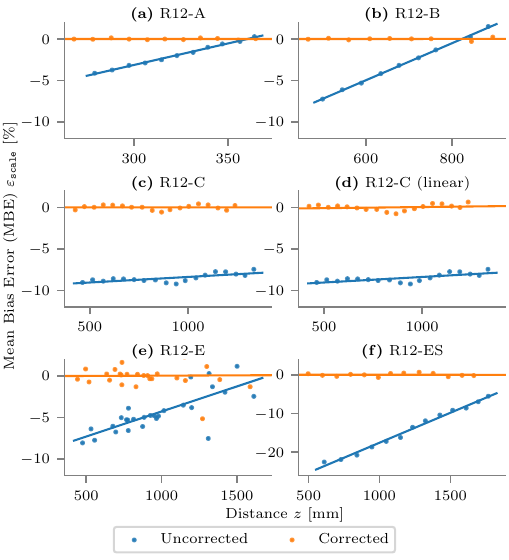}
	\caption{
		Scale errors before and after correction as function of the distance for the datasets \texttt{R12-A} \textbf{(a)}, \texttt{R12-B} \textbf{(b)}, \texttt{R12-C} \textbf{(c)}, \texttt{R12-E} \textbf{(e)} and \texttt{R12-ES} \textbf{(f)} with their associated fitting functions. 	
		A positive error means the object is smaller than the reference, and a negative error means that the objects is bigger than the reference.
		The absolute error grows when getting farther from the focus distance. 
		The corrected distance has a nearly null error.
		Sub-figures \textbf{(c)} and \textbf{(d)} give a comparison of the linear model versus the quadratic model scale correction on the dataset \texttt{R12-C}. 
		The quadratic model performs slightly better than the linear model, as the slope of the fitted line is closer to zero compared to the slope of the linear fitted model.
	}\label{fig:scalerr}\label{fig:scalecomp}%
\end{figure}

\noindent\textbf{Depth scaling correction.}
Coarse depth estimation for datasets \texttt{R12-A,B,C,ES} is performed on images corresponding to planar checkerboards orthogonal to the optical axis and uniformly distributed in the range of distances, whilst for \texttt{R12-E}, depth estimation is performed on hand-held checkerboards, leading to noisier evaluation of depth estimates.
The reported depth associated to each frame is the median of the depth estimates. 
Calibrated depth scaling coefficients are reported in \autoref{tab:scalecoeff}, along with their median scale error after correction for the evaluation datasets.
Depths before and after correction are illustrated in \autoref{fig:scalerr}.
A positive error means the estimated object is smaller than the ground truth, and a negative error means that the estimated object is bigger than the ground truth.
The absolute error grows when getting farther from the focus distance. 
All corrected distances have a nearly null scale error.
For all datasets, {with real data from four configurations and also in simulation}, our methodology successfully corrects the scale, with a final median scale error $\fun{\varepsilon_{scale}}$ of less than \SI{0.05}{\percent}, {indicating thus nearly no scale error}. \\ %

\noindent\textbf{Analysis on simulated data.} We investigated scale error on simulated images of the \texttt{R12-ES} dataset. 
Depths are estimated based on the coarse depth estimation framework, for ground truth distances from \SIrange{500}{1900}{\mm} with a step of \SI{100}{\mm}.
Without scale correction, depths are estimated from \SIrange{610.9}{1917.4}{\mm} with a mean relative error $\varepsilon_z = 8.06~\si{\percent}$.
After scale correction, depths are {closer to the ground truth and are} estimated from \SIrange{495.7}{1841.8}{\mm} with a mean relative error reduced by a factor two, $\varepsilon_z = 3.78~\si{\percent}$.
As shown by \autoref{fig:scalerr}, depth scaling error appears even in simulated data, showing that this phenomenon must be added to the inverse projection model to reach precise and accurate depth measurements.\\

\noindent\textbf{Scaling model comparison.} Secondly, we evaluated the choice of the {empirical} scaling model and presented the results for the dataset \texttt{R12-C}. As illustrated in \autoref{fig:scalecomp}, the quadratic model performs slightly better than the linear model, as the slope of the fitted line is closer to zero compared to the slope of the linear fitted model.
This is confirmed by the median scale error after correction reported in \autoref{tab:scalecoeff} which is reduced by a factor two with the quadratic model.
In the following, depths will be corrected with the quadratic model.

\begin{figure*}[!t]
	\includegraphics[width=\linewidth]{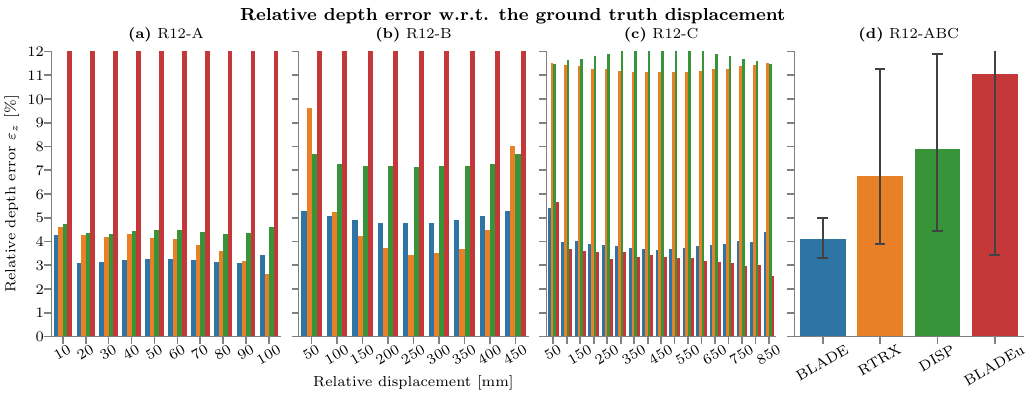}
	\caption{
		Relative depth error along the $z$-axis with respect to the ground truth displacement from the closest frame, for datasets \texttt{R12-A} \textbf{(a)}, \texttt{R12-B} \textbf{(b)} and \texttt{R12-C} \textbf{(c)}.
		The error $\varepsilon_z$ is expressed in percentage of the estimated distance, and truncated to \SI{12}{\percent} to ease the readability and the comparison.
		The mean error with its confidence interval across all datasets for our method with (\texttt{BLADE}) and without scale correction (\texttt{BLADEu}), for the model of \cite{Noury2017b} using only disparity (\texttt{DISP}), and for the proprietary software \texttt{RxLive} (\texttt{RTRX}) are reported in \textbf{(d)}.
		Please refer to the color version for better visualization.
	}\label{fig:relativedeptherrors}
\end{figure*}

\begin{table}[!t]
	\caption{Depth scaling coefficients for datasets \texttt{R12-A,B,C} and \texttt{R12-E,ES} with the median scale error after correction. For dataset \texttt{R12-C}, the error for the linear model is also reported.}\label{tab:scalecoeff}%
	\centering\footnotesize%
	\begin{tabu}{l XXX r}
		\toprule
		& $\gamma_2\left(\times10^{-5}\right)$ & $\gamma_1$ & $\gamma_0$ & $\fun{\varepsilon_{scale}} \left(\si{\percent}\right)$ \\
		\midrule
		\midrule
		\texttt{R12-A} & \nump{3}{79.708898900074192}& \nump{3}{0.62515220840000596} & \nump{3}{31.51235516605901}& \nump{3}{-0.0135038}\\
		\texttt{R12-B} & \nump{3}{23.017216184680102}& \nump{3}{0.79609141585669219}& \nump{3}{10.912844125074178}& \nump{3}{0.0224473}\\
		\texttt{R12-C} & \nump{3}{2.7355268379573649} & \nump{3}{0.88338184462070446} & \nump{3}{10.90983152629782}& \nump{3}{0.0238045}\\
		\texttt{R12-C} & - & \nump{3}{0.92900210913956494} & \nump{3}{-6.1013017162389138} & \nump{3}{0.0434072}\\
		\midrule
		\texttt{R12-ES} & \nump{3}{14.430517978719409} & \nump{3}{0.66663820764239778} & \nump{3}{35.910141602595473} & \nump{3}{-0.0230534}\\
		\texttt{R12-E} & \nump{3}{4.6169969732325145}& \nump{3}{0.91160914393108972} & \nump{3}{-2.004049745583496} & \nump{3}{0.0465172}\\
		\bottomrule
	\end{tabu}
\end{table}

\subsection{Relative depth evaluation}

We compared our method with (\texttt{BLADE}) and without (\texttt{BLADEu}) depth scaling on relative depth estimation with state-of-the-art methods, including the \texttt{Raytrix} software \citep{Perwass2010c}, corresponding to the model of \citet{Heinze2016b} (denoted  \texttt{RTRX}), and depth estimation with the model of \citet{Noury2017b} using only the disparity (denoted \texttt{DISP}) in our framework. The latter provides a baseline using a classic block matching technique but that does not leverage blur information.
The relative depth errors along the $z$-axis with respect to the ground truth displacement from the closest frame are reported in \autoref{fig:relativedeptherrors} for datasets \texttt{R12-A} {(a)}, \texttt{R12-B} {(b)} and \texttt{R12-C} {(c)}.
The mean error with its confidence interval across all datasets is illustrated in {(d)} for each method, and is: 
for \texttt{BLADE}, $\varepsilon_z = 4.09 \pm 0.85~\si{\percent}$;
for \texttt{BLADEu}, $\varepsilon_z = 11.05 \pm 6.61~\si{\percent}$;
for \texttt{DISP}, $\varepsilon_z = 7.88 \pm 3.74~\si{\percent}$; and,
for \texttt{RTRX}, $\varepsilon_z = 6.76 \pm 3.96~\si{\percent}$.
First, we see that our scale correction effectively improves the depth estimation results {, as expected from our depth scaling error analysis}.
With scale correction it outperforms the other methods, {meaning that leveraging both defocus and correspondences provides accurate results.}
This behavior is also validated in simulation on \texttt{R12-ES}, where the mean error after correction is of the same order, \ie $\varepsilon_z = 3.78~\si{\percent}$.

Secondly, the \texttt{BLADE} method presents the lowest standard deviation, a stable error across all distances, and errors of the same order for all configurations. 
The other methods vary significantly across the datasets, making our method the only one able to generalize to several configurations without losing precision. {It thus means that leveraging both defocus and correspondences provides precise results.}

Finally, for datasets \texttt{R12-A,B,ES}, the scale correction clearly improves the relative depth estimates.
In \texttt{R12-C}, the results are similar with and without. 
The errors are nearly constant for the evaluated range when uncorrected as illustrated in \autoref{fig:scalerr}.
The effect of scale correction is less visible as it can be assimilated to a bias correction.

\subsection{Absolute depth evaluation on 3D scenes}

We used the dataset \texttt{R12-ELP20} to evaluate our depth estimation framework on absolute metric depth estimates.\\

\noindent\textbf{Central sub-aperture depth map.} To compare depth estimates, we generated for each scene the \ac{CSAD} as follows: 
\begin{enumerate}
\item From the raw depth map, we back-project each pixel having a virtual depth hypothesis into a 3D point in metric space;
\item We replace the plenoptic camera model by a pinhole model, where the sensor is now at a distance $\Focal$ from the main lens, and we increase the pixel size by a factor $S$ (here, $S=4$, the final resolution is thus $1020 \times 767~\si{pixels}$);
\item We project each point of the point cloud with the new pinhole model, using a $z$-buffer like technique, and attributing the minimum depth value to the pixel (or the median value for noisy data);
\item And finally, we filter the resulting depth map image by applying a median filter and a morphological erosion to reduce noise.
\end{enumerate} 
For generating ground truth \acp{CSAD}, we replace the first step by simply applying the extrinsic transformation ${}^{c}{\Transform}_{{l}}$ so that the 3D points are expressed in camera frame.\\
\noindent\textbf{Evaluated methods.} We evaluated our \texttt{BLADE} framework considering the following variations: 1) using relative blur information (\texttt{B}) or only disparity (\texttt{D}); 2) using the coarse (\texttt{C}) or the refined (\texttt{R}) estimation; and 3) using the scale-corrected (\texttt{S}) or scale-uncorrected (\texttt{U}) model.
Note that we use the same intrinsic parameters for all evaluations.
In the end, we presented the results for eight methods.
For each method, we generated \ac{CSAD} for each of the five scenes of \texttt{R12-ELP20}, and compared them to the ground truths. {It allows us to validate both the scaling model, and the improvement of accuracy due to the defocus information.}\\

\noindent\textbf{Metrics.} To analyze the depth error, we computed a quality map as the absolute difference (AD) between the depth map and the ground truth.
As the maps are sparse and not dense, we generated a mask corresponding to pixels in common where depth estimates are available.
Errors are thus computed only for pixels in the mask.
Finally, statistics over the errors are computed.
We used the percentiles (at \num{25} and \SI{75}{\percent}) and the median to describe the overall error of the depth map estimates. \\

\begin{table*}[!t]
\caption{
	Statistics (percentiles and median) of the absolute difference (AD) error of central sub-aperture depth map for each variation of our \texttt{BLADE} framework (using relative blur information (\texttt{B}) or only disparity (\texttt{D}); using the coarse (\texttt{C}) or the refined (\texttt{R}) estimation; and using the scale-corrected (\texttt{S}) or scale-uncorrected (\texttt{U}) model), on the scenes of dataset \texttt{R12-ELP20}.
	All errors are expressed in \si{\mm}. 
	The last column indicates the mean of the median errors for all scenes.
}\label{tab:3dscenedeptherrors}
\footnotesize
\begin{tabu}{ccc XXX XXX XXX XXX XXX c}
	\toprule
	\multirow{2}{*}{\rotatebox[origin=r]{90}{\texttt{B/D}}} &
	\multirow{2}{*}{\rotatebox[origin=r]{90}{\texttt{C/R}}} &
	\multirow{2}{*}{\rotatebox[origin=r]{90}{\texttt{S/U}}} &
	\multicolumn{3}{c}{\texttt{Calib}}& \multicolumn{3}{c}{\texttt{Plane-1}} & \multicolumn{3}{c}{\texttt{Plane-2}}& \multicolumn{3}{c}{\texttt{Figurines-1}} & \multicolumn{3}{c}{\texttt{Figurines-2}}&
	\multirow{2}{*}{Total}  \\
	\cmidrule(lr){4-6}\cmidrule(lr){7-9}\cmidrule(lr){10-12}\cmidrule(lr){13-15}\cmidrule(lr){16-18}
	&&& $\mathcal{Q}_{25}$ & med. & $\mathcal{Q}_{75}$ & $\mathcal{Q}_{25}$ & med. & $\mathcal{Q}_{75}$& $\mathcal{Q}_{25}$ & med. & $\mathcal{Q}_{75}$& $\mathcal{Q}_{25}$ & med. & $\mathcal{Q}_{75}$& $\mathcal{Q}_{25}$ & med. & $\mathcal{Q}_{75}$& \\
	\midrule
	\midrule 
	\texttt{B} & \texttt{C} & \texttt{S} %
	& \textbf{7.661} & \textbf{17.761} & \textbf{39.146} %
	& \textbf{6.498} & \textbf{13.940} & \textbf{26.402}%
	& \textbf{5.482} & \textbf{11.370} & \textbf{24.468}%
	& \textbf{11.720} & \textbf{24.867} & \textbf{47.270} %
	& \textbf{12.310} & \nump{3}{27.0575} & \nump{3}{54.395} %
	& \textbf{18.999} \\
	\texttt{D} & \texttt{C} & \texttt{S} %
	& \nump{3}{7.9162} & \nump{3}{18.6476} & \nump{3}{39.1555} %
	& \nump{3}{7.0022} & \nump{3}{14.6721} & \nump{3}{27.5994} %
	& \nump{3}{5.64429} & \nump{3}{12.0071} & \nump{3}{25.0219} %
	& \nump{3}{12.4457} & \nump{3}{26.3074} & \nump{3}{50.5537} %
	& \nump{3}{13.5137} & \nump{3}{29.8092} & \nump{3}{57.6658} %
	& \nump{3}{20.28868} \\
	\texttt{B} & \texttt{R} & \texttt{S} %
	& \nump{3}{8.83105} & \nump{3}{19.9607} & \nump{3}{43.8795} %
	& \nump{3}{11.7651} & \nump{3}{21.8179} & \nump{3}{34.9033} %
	& \nump{3}{9.75812} & \nump{3}{17.9507} & \nump{3}{28.9246} %
	& \nump{3}{12.0111} & \nump{3}{25.2824} & \nump{3}{50.3767} %
	& \nump{3}{14.2029} & \nump{3}{29.4424} & \nump{3}{56.8995} %
	& \nump{3}{22.89082} \\
	\texttt{D} & \texttt{R} & \texttt{S} %
	& \nump{3}{9.73602} & \nump{3}{21.9922} & \nump{3}{49.8718} %
	& \nump{3}{14.7998} & \nump{3}{25.9039} & \nump{3}{39.2953} %
	& \nump{3}{11.0743} & \nump{3}{20.6133} & \nump{3}{32.0541} %
	& \nump{3}{13.5514} & \nump{3}{28.0156} & \nump{3}{55.308} %
	& \nump{3}{16.9114} & \nump{3}{34.1245} & \nump{3}{62.5741} %
	& \nump{3}{26.1299} \\
	\midrule
	\texttt{B} & \texttt{C} & \texttt{U} %
	& \nump{3}{34.45} & \nump{3}{50.1063} & \nump{3}{68.9974} %
	& \nump{3}{30.7433} & \nump{3}{42.452} & \nump{3}{58.2409} %
	& \nump{3}{33.657} & \nump{3}{42.9047} & \nump{3}{55.6637} %
	& \nump{3}{21.4171} & \nump{3}{43.5767} & \nump{3}{67.5707} %
	& \nump{3}{18.415} & \nump{3}{37.834} & \nump{3}{63.5215} %
	& \nump{3}{43,37474} \\
	\texttt{D} & \texttt{C} & \texttt{U} 
	& \nump{3}{31.1049} & \nump{3}{47.0714} & \nump{3}{66.9825} %
	& \nump{3}{27.5806} & \nump{3}{39.3578} & \nump{3}{54.3869} %
	& \nump{3}{30.7535} & \nump{3}{40.0814} & \nump{3}{51.9254} %
	& \nump{3}{18.3254} & \nump{3}{40.3228} & \nump{3}{63.9747} %
	& \nump{3}{15.7125} & \nump{3}{33.3715} & \nump{3}{58.125} %
	& \nump{3}{40.04098} \\
	\texttt{B} & \texttt{R} & \texttt{U} %
	& \nump{3}{27.1142} & \nump{3}{41.3267} & \nump{3}{63.5831} %
	& \nump{3}{17.5525} & \nump{3}{28.3035} & \nump{3}{40.732} %
	& \nump{3}{20.0247} & \nump{3}{30.1921} & \nump{3}{39.835} %
	& \nump{3}{16.5389} & \nump{3}{30.2432} & \nump{3}{52.072} %
	& \nump{3}{14.3218} & \nump{3}{28.8776} & \nump{3}{48.5909}%
	& \nump{3}{31.78862} \\
	\texttt{D} & \texttt{R} & \texttt{U} %
	& \nump{3}{26.2738} & \nump{3}{41.5221} & \nump{3}{64.9519} %
	& \nump{3}{14.0977} & \nump{3}{24.2548} & \nump{3}{37.0621} %
	& \nump{3}{17.1154} & \nump{3}{27.317} & \nump{3}{38.432} %
	& \nump{3}{15.2722} & \nump{3}{28.6899} & \nump{3}{50.1523} %
	& \nump{3}{12.787} & \textbf{26.733} & \textbf{46.932} %
	& \nump{3}{29.7033} \\
	\bottomrule
\end{tabu}		
\end{table*}

\begin{figure*}[!t]
\centering
\includegraphics[width=\linewidth]{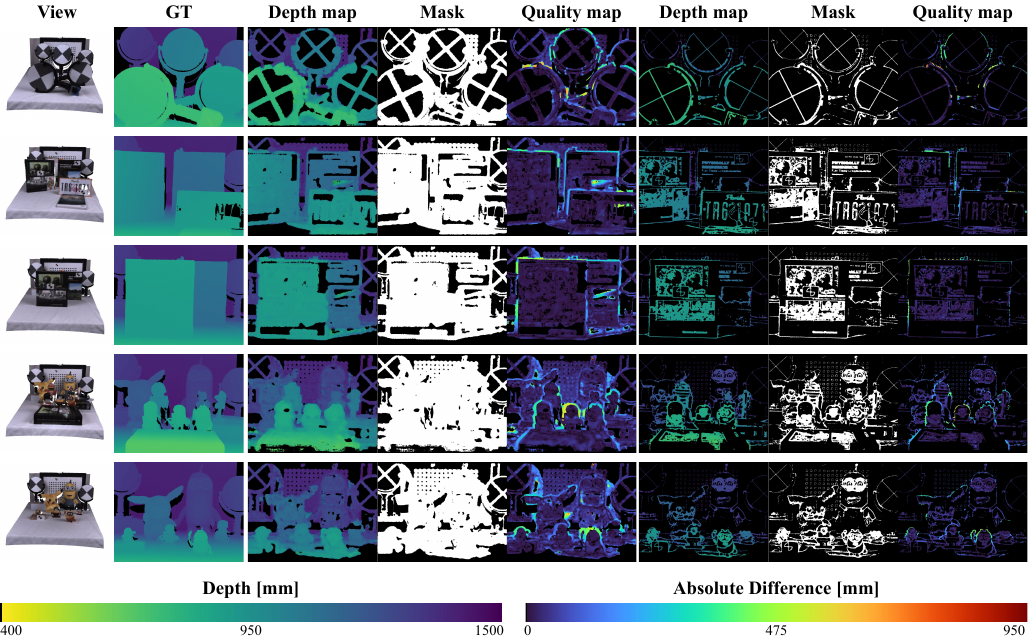}
\caption{
	Snapshot view of the colored point cloud, along with the ground truth \acf{CSAD} for each scene of the dataset \texttt{R12-ELP20}.
	\ac{CSAD}, mask and quality map representing the absolute difference (AD) error are illustrated for the coarse and refined variations (\ie \texttt{B/C/S} and \texttt{B/R/S}) of our framework. 
	Results for the other variations can be found in supplementary materials.
	Please refer to the color version for better visualization.
}\label{fig:3dscenedeptherrors}
\end{figure*}

\noindent\textbf{Results.} 
Statistics for all the variations of the \texttt{BLADE} framework for all the scenes are reported in \autoref{tab:3dscenedeptherrors}. 
Bold font indicates the best results. 
The last column reports the mean of the median errors for all scenes. 
Snapshot of the colored point cloud, along with the ground truth \ac{CSAD}, are reported for each scene in \autoref{fig:3dscenedeptherrors}.
Depth map, mask and quality map are illustrated for the coarse and refined variations (\ie \texttt{B/C/S} and \texttt{B/R/S}) of our framework. 
Results for the other variations can be found in supplementary material.

From the reported errors, the scenes can be divided into two groups: 1) the easy scenes (\texttt{Calib}, \texttt{Plane-1} and \texttt{Plane-2}) containing mostly planar objects and presenting the lowest errors; and 2) the complex scenes (\texttt{Figurines-1} and \texttt{Figurines-2}) containing more objects with less texture and more complex shapes, presenting a larger error.
The errors distributions are similar for all the scenes, {i.e., most of the errors are located on the objects' border.}

First of all, the lower overall error is obtained for the variation leveraging blur in our coarse depth estimation framework (\texttt{B/C/S}).
The mean median-error over all scenes is less than \SI{19}{\mm}, for distances ranging from \SIrange{400}{1500}{\mm}.
It corresponds to relative errors ranging from \SIrange{1.27}{4.75}{\percent} of the distance, which is coherent with the relative depth evaluation.
For easy scenes, relative errors range from \SIrange{0.96}{3.59}{\percent} of the distance.
For complex scenes, relative errors range from \SIrange{1.73}{6.49}{\percent} of the distance.
As illustrated in the quality maps, most of the errors are located at objects boundaries, whereas the errors are low everywhere else. 
This is due to our method not explicitly dealing with occlusion boundaries, leading to wrong estimates in those regions.
Second, it is clear that the scaled variations outperform the unscaled ones, {as expected from the previous experiments.}
Our depth scaling calibration efficiently corrects the depth estimates, and allows to generate metric depth map without scale errors.
Third, integrating both correspondence and defocus cues shows lower median error and lower percentiles for all scenes, compared to only using disparity. {It is due to compensating the mismatch of focus, which both helps the matching process and provides more cues about depth to be exploited.}
Finally, coarse estimation leads to denser depth maps as illustrated in \autoref{fig:3dscenedeptherrors}, and is able to extrapolate depth where there is enough texture within the micro-image using a locally planar approximation.
Refined estimation captures depth information only on the textured areas.
The maps are sparser but wrong estimates at object boundaries do not spread as much as using the locally planar approximation. {Since proportionally there is more pixels corresponding to borders, the error is a bit higher for the refined version. With better strategy to deal with occlusion, results are expected to be similar for both. The choice of using which variation depends on the application behind: if we need more details, refined is preferred; if a denser map is required, the coarse is preferred.}\\
Note that for the scene \texttt{Figurines-2}, the variation \texttt{D/R/U} has a smaller median error than the others.
Recall that for \texttt{R12-E}, objects are reconstructed farther when unscaled (\ie the scale error is negative, see \autoref{fig:scalerr}).
As most of the errors appear on the objects' boundaries, objects in foreground are closer to the reference ones in background, and the difference of depth is then smaller.
Furthermore, in the refined case, most of the estimations are done only on locally textured areas such as those boundaries.
Combination of those two factors allows to explain why this variation has a smaller error.
\section{Conclusion}\label{sec:ccl}

With a plenoptic camera, depth estimation and 3D reconstruction can be performed directly from a single acquisition, with scale information.
Inherently from its design, a multi-focus plenoptic camera captures both correspondence and defocus cues which are complementary for depth estimation.

In this paper, we presented a new metric depth estimation algorithm using only raw images from {multi-focus} plenoptic cameras.
It is especially suited for the multi-focus configuration where several micro-lenses with different focal lengths are used.
First, we completed our previous camera model by introducing the inverse projection model.
We showed that depth recovered from virtual depth hypothesis suffers from a scale error.
We included then an {empirical} depth scaling correction model as well as a methodology to calibrate it.
Second, we introduced our blur aware depth estimation (BLADE) framework, improving disparity estimation for defocus stereo images via compensating the mismatch of focus, \ie integrating both correspondence and defocus cues.
We formulated then a new residual error to leverage blur information for depth estimation which is used in two variations of our framework to recover depth either per micro-image or per pixel.
Finally, our results showed that introducing defocus cue improves the depth estimation.
We demonstrated the effectiveness of our depth scaling calibration on relative depth estimation setup and on real-world 3D complex scenes with ground truths acquired with a 3D lidar scanner.
With our method, we obtained a median relative depth error ranging from \SIrange{1.27}{4.75}{\percent} of the distance. 
In our experiments it corresponds to a median error of less than \SI{19}{\mm}, for distances ranging from \SIrange{400}{1500}{\mm}.\\

\noindent\textbf{Discussions on limitations and improvements.}
{Regarding the generalization of our method to other plenoptic cameras, i.e., simple focused or unfocused plenoptic cameras, the same camera model can be exploited, though no defocus cues are included in the depth estimation process. Thanks to the scale correction model, metric precise and accurate depth should be expected.} \\
{One limitation of our method is that} computational cost is not addressed here. 
Our method is implemented purely on CPU with a brute force algorithm for finding the minimum of the cost function  {(about 10 to 30 seconds per frame)}. 
We can leverage neighborhood information and implement a belief propagation strategy to avoid having to compute an initial hypothesis for each micro-image. 
Combined with a GPU implementation, computation time can be significantly improved.
Furthermore, as most of the errors are located at object boundaries, we can adapt several strategies: to explicitly manage occlusions; to check coherence between estimates ref-target and target-ref; to model uncertainty as the weight function in the cost function; and we can proceed to a robust filtering to eliminate outliers. %
Global refinement can also be considered as further steps to improve depth estimates.
Future work will include the discussed improvements, as well as using the depth map for metrology and robotic applications.
\section*{Acknowledgments}

This work was supported by the AURA Region and the European Union (FEDER) through the MMII project of CPER 2015-2020 MMaSyF challenge.
We would like to thank Laurent Malaterre, Trong-Lanh R\'emi Nguyen, Ruddy Theodose and Etienne Cadet for their help during the acquisitions.
{We also thank the reviewers for their insightful comments.}

{
	\bibliographystyle{model2-names}
	\interlinepenalty=10000
	\bibliography{manuscript}
}

\cleardoublepage
\appendix

\section*{\huge Supplementary Material}

As supplementary material, we 
first provide details about the projection model regarding the distances $\funarg{\distmlasensor}{\mlindexes}$ (\ie the distance between the micro-lens and the sensor) and $\funarg{\distlensmla}{\mlindexes}$ (\ie the distance between the micro-lens and the main lens) for each micro-lens $\left(\mlindexes\right)$.
Second, we comment on the depth scaling phenomenon.
Third, we quantify the approximation used in Eq.~18.
For completeness, we then report the intrinsic and extrinsic parameters for the camera setup used in the evaluation.
Finally, we give the additional results for corrected disparity analysis and for all variations of the method for the absolute depth evaluation on 3D scenes.

\section{Details on direct projection model}

As noted in \cite{Labussiere2022calib}, technically we should consider specific distances $\funarg{\distmlasensor}{\mlindexes}$ (\ie the distance between the micro-lens and the sensor) and $\funarg{\distlensmla}{\mlindexes}$ (\ie the distance between the micro-lens and the main lens) for each micro-lens $\left(\mlindexes\right)$. \\

It corresponds to take the \ac{MLA} tilt into consideration in the projection model.
In this paper, we improve our model by taking into account the effect of the \ac{MLA} tilt with respect to the sensor. 
It is reduced by using an orthogonal approximation and thus considering specific distances $\funarg{\distmlasensor}{\mlindexes}$ (\ie the distance between the micro-lens and the sensor) and $\funarg{\distlensmla}{\mlindexes}$ (\ie the distance between the micro-lens and the main lens) for each micro-lens $\left(\mlindexes\right)$.
To ease the reading, we only used the notation $\distlensmla$ and $\distmlasensor$, but the quantities can be replaced by their corresponding orthogonal approximation.

\section{Comments on depth scaling}

At the end of the BLADE process, we have a virtual depth estimate associated to each pixel.
To obtain a metric depth, we use the inverse projection model given in Eq.~5, where $\blurradiuspix$ is computed from the virtual depth by Eq.~4, \ie 
\begin{equation*}
\blurradiuspix = m \cdot \virtdepth^{-1} + q_i'\tdot
\end{equation*}

\begin{figure}[!t]
	\centering
	\includegraphics[width=\linewidth]{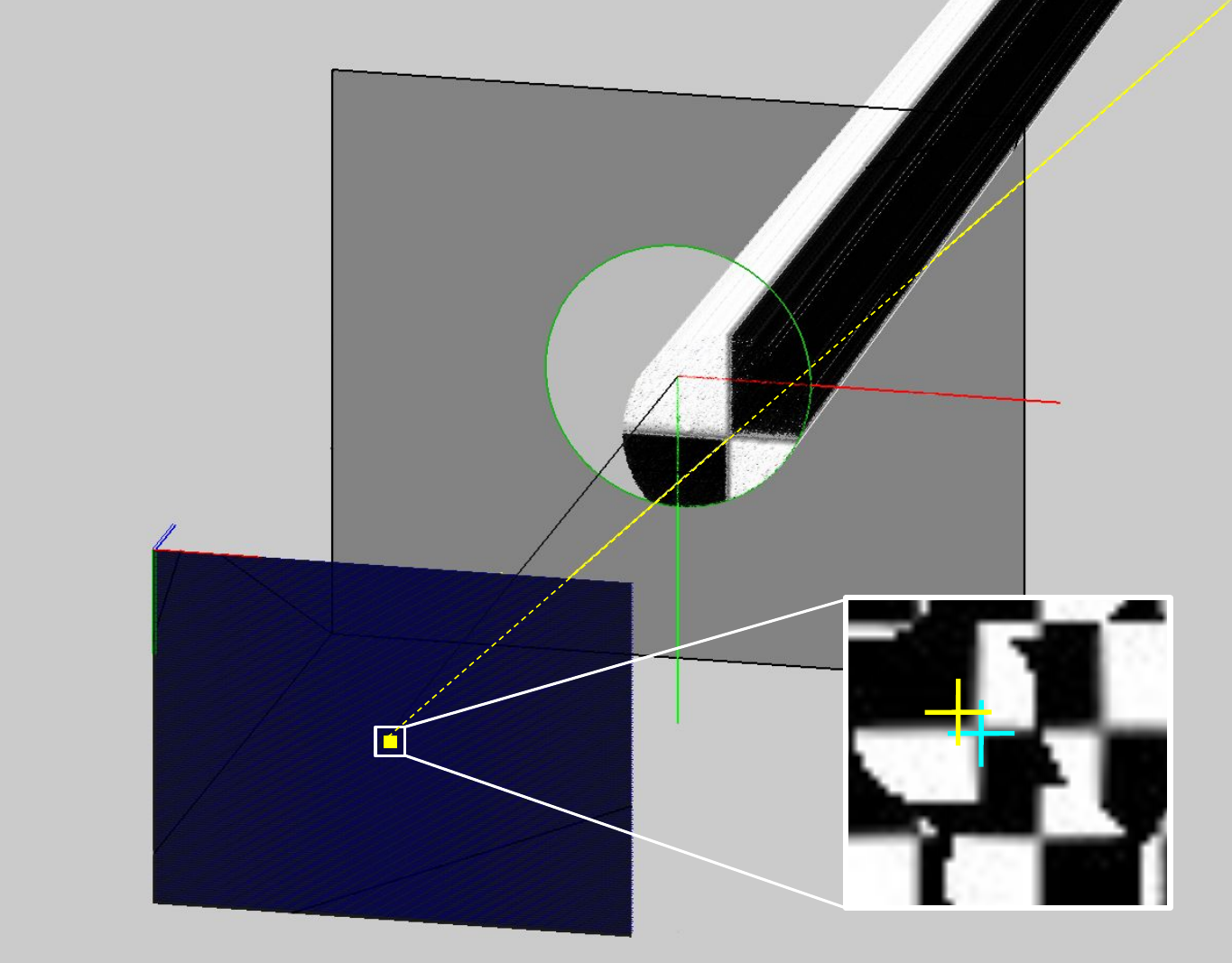}
	\caption{Illustration of the ray-tracing from a pixel containing a checkerboard corner with a real aperture. All elements are illustrated at the same scale (except the pixel size). Blue plane represents the \ac{MLA} and the sensor planes. The gray plane is the main lens plane. The green circle models the real aperture. The yellow ray is the chief ray emanating from the considered pixel and going through the micro-lens center. Black and white rays are the rays from the cone of light that reach the pixel. As illustrated, the chief ray is correctly hitting the checkerboard corner in object space. However, since not all the rays from the cone of light go through the aperture, integration of the radiance from the passing ones induces a shift in the observed radiance in sensor space.
		Indeed, the theoretical projection of the corner in yellow corresponding to the yellow chief ray is not matching the observed corner position in cyan.
	}
	\label{fig:subapchiefray}
\end{figure}

The algorithm effectively retrieves the virtual depth hypothesis corresponding to the observed images, \ie the relation 
\begin{equation}
\hat{\virtdepth} = \frac{\baseline}{\baseline - \Delta\!\vect{p}}
\end{equation}
is verified, where $\Delta\!\vect{p}$ is the Euclidean distance between two observations. 
It means that the correct disparity is correctly estimate from image space.
However, when mapping from the virtual space to the object space with the inverse projection model, we observe that reconstructed objects are scaled up to a certain factor, in the $z$-dimension (\eg points are projected farther) but also in the $xy$-dimensions (\eg objects appear bigger). 
The scale factor grows approximately linearly as function of the distance with respect to the focus distance (see Fig.~7).
This phenomenon appears both on real and simulated data.

It is due to the limitations of the thin-lens model.
The proposed model describes efficiently the projective geometry for a point whose all rays inside the projection cone attain the retina.
Indeed, when simulating the imaging process with a large aperture, the retrieved depths from the simulated images do not suffer from a scale error.
All rays can go through, and the final radiance is the correct sum of all rays contained within the cone.
With real aperture, some rays, including the chief ray, emanating from the pixel do not always go thought the main lens and are blocked, as shown in \autoref{fig:subapchiefray}.
The observed radiance should be the integrate of all rays within the cone but only a portion of it can go through the aperture, meaning that the radiance associated to the projection is not the one that is effectively observed.
This phenomenon must thus be taken into account.

\section{Quantification of the approximation in Eq.~18}

Let $a=2 n d$ be the distance to the micro-lens $\left(i\right)$, with $n \in \range{1,10}$, such as $\virtdepth = a / d = 2n$. 
Let $\delta\!z = n B \sin\!\left(\alpha\right) $ be the $z$-shift between the micro-lenses $\left(i\right)$ and $\left(j\right)$ separated by a distance $nB$. 
The virtual depth $v'$ for the micro-lens $\left(j\right)$ is then given by $\virtdepth' = \left(a+\delta\!z\right) / \left(d+\delta\!z\right)$.\\	
First, as $B < d / 2$, and with $\sin\!\left(\alpha\right) \approx \alpha$ since $\alpha$ is small, we can can expressed $v'$ as 
\begin{equation}
\virtdepth'  = \frac{a+\delta\!z}{d+\delta\!z}
= \frac{a + \alpha nB}{d +\alpha nB}
< \frac{2nd + \alpha\frac{n}{2}d}{d + \alpha\frac{n}{2}d}
< \frac{2 + \frac{\alpha}{2}}{\frac{1}{n} + \frac{\alpha}{2}}\tdot
\end{equation}
Second, let measures the approximation error of $\virtdepth'$ by $\virtdepth$. The relative error is given by 
\begin{equation}
\begin{aligned}
\epsilon &= \abs{1- \frac{\virtdepth}{\virtdepth'}}
= \abs{1 - 2n \cdot \frac{1/n + \alpha/2}{2 + \alpha/2}} \\
&= \abs{1 - \frac{2 + n\alpha}{2 + \alpha/2}}
= \abs{\frac{\alpha/2 - n\alpha}{2 + \alpha/2}}\tdot
\end{aligned}
\end{equation}
Finally, from calibration, $\alpha < 0.0005~\si{rad}$, the relative error is thus bounded such that $0.0125~\% < \epsilon_\% < 0.2375~\%$, which makes the approximation valid.
\qed
\section{Intrinsic and extrinsic parameters}

Intrinsic parameters for datasets \texttt{R12-A,B,C,E} are reported in \autoref{tab:intrinsics} for our camera model \cite{Labussiere2020blur} (\texttt{BAP}), for the model of \cite{Noury2017b} (\texttt{NOUR}), and for the proprietary \texttt{RxLive} software of \texttt{Raytrix} corresponding to the model of \cite{Heinze2016b} (\texttt{RTRX}). \\

The extrinsic parameters correspond to the pose between the camera and the 3D lidar scanner.
In our experiments, using a constellation of five points, the obtained optimized transformation is
\begin{equation*}\label{eq:extrinsics}
{}^{c}{\Transform}_{{l}} = \bbm  -0.88925& -0.070239&   0.451992&177.278 \\
-0.456949&  0.0917283&  -0.88475& -394.36 \\
0.020684&  -0.993304&  -0.11366& -266.858 \\ 
0&0&0&1 \ebm\tdot 
\end{equation*}

\section{Complementary results}

\subsection{On corrected disparity}
We evaluated the impact of the orthogonal approximation of the micro-lens baseline with respect to the micro-image baseline, \ie using the corrected disparity (corresponding to Eq.~(29)) instead of the commonly used disparity formulation (corresponding to Eq.~(28)).
Analysis is performed on dataset \texttt{R12-A}.
Without depth scaling correction, we have for the approximated disparity a mean relative error $\varepsilon_z = 19.38~\si{\percent}$, which is reduced to $\varepsilon_z = 14.99~\si{\percent}$ only by using the corrected disparity formulation.
We thus used the disparity obtained from Eq.~(29) for all evaluations.

\subsection{On depth evaluation on 3D scenes}

Recall that we evaluated our \texttt{BLADE} framework considering the following variations: 1) using relative blur information (\texttt{B}) or only disparity (\texttt{D}); 2) using the coarse (\texttt{C}) or the refined (\texttt{R}) estimation; and 3) using the scale-corrected (\texttt{S}) or scale-uncorrected (\texttt{U}) model.

\begin{figure*}[!t]
	\centering
	\includegraphics[width=\linewidth]{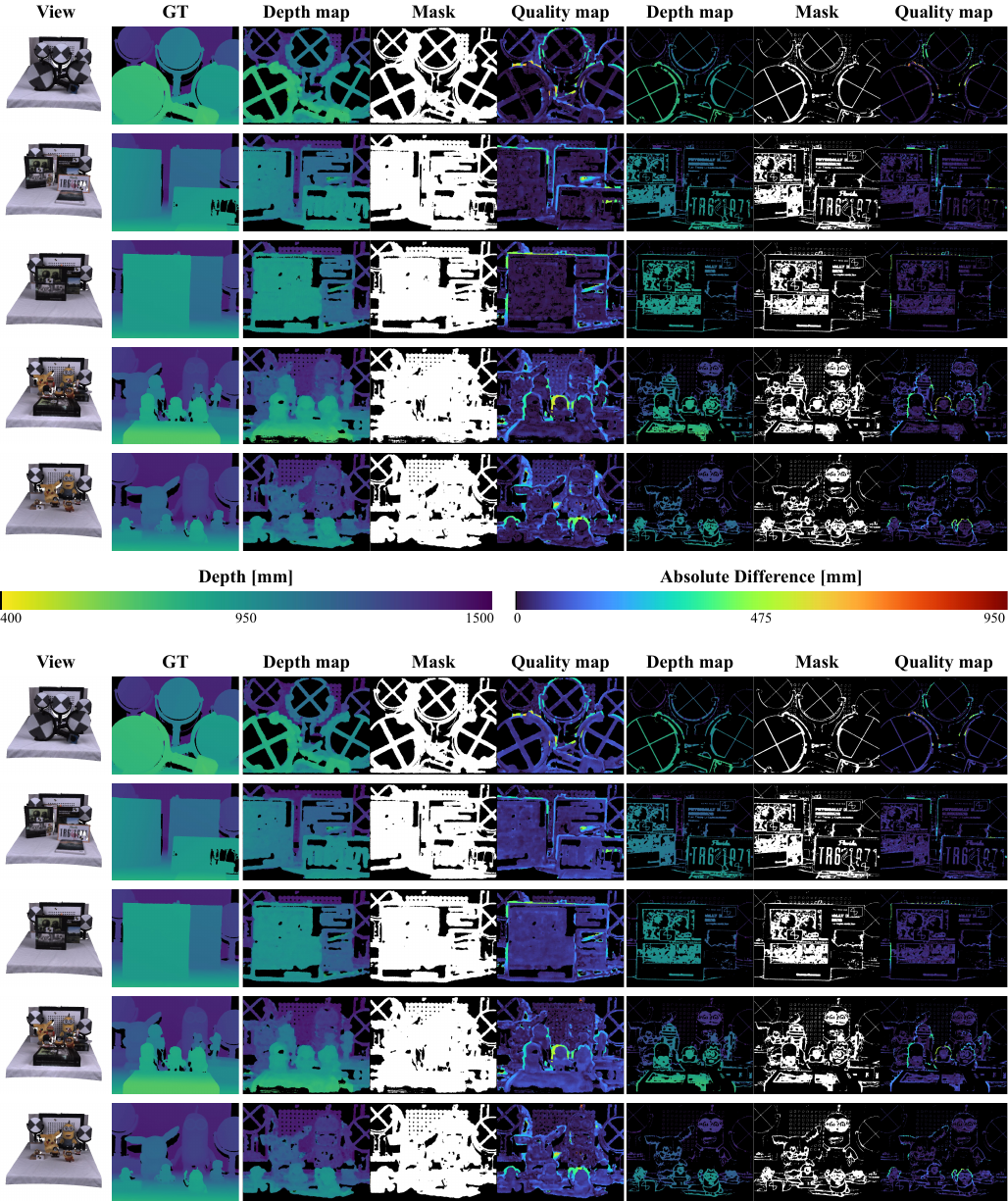}
	\caption{
		Snapshot view of the colored point cloud, along with the ground truth \acf{CSAD} are reported for each scene of the dataset \texttt{R12-ELP20}.
		\ac{CSAD}, mask and quality map representing the absolute difference (AD) error are illustrated for all variations using the relative blur (\texttt{B}) of our framework:
		\textit{top} is  the scaled (\texttt{S}) coarse (\texttt{C}) and refined variations (\texttt{R}), 
		\textit{bottom} is the unscaled (\texttt{U}) coarse (\texttt{C}) and refined variations (\texttt{R}).
		Please refer to the color version for better visualization.
	}\label{fig:3dscenedeptherrors-blade}
\end{figure*}

\begin{figure*}[!t]
	\centering
	\includegraphics[width=\linewidth]{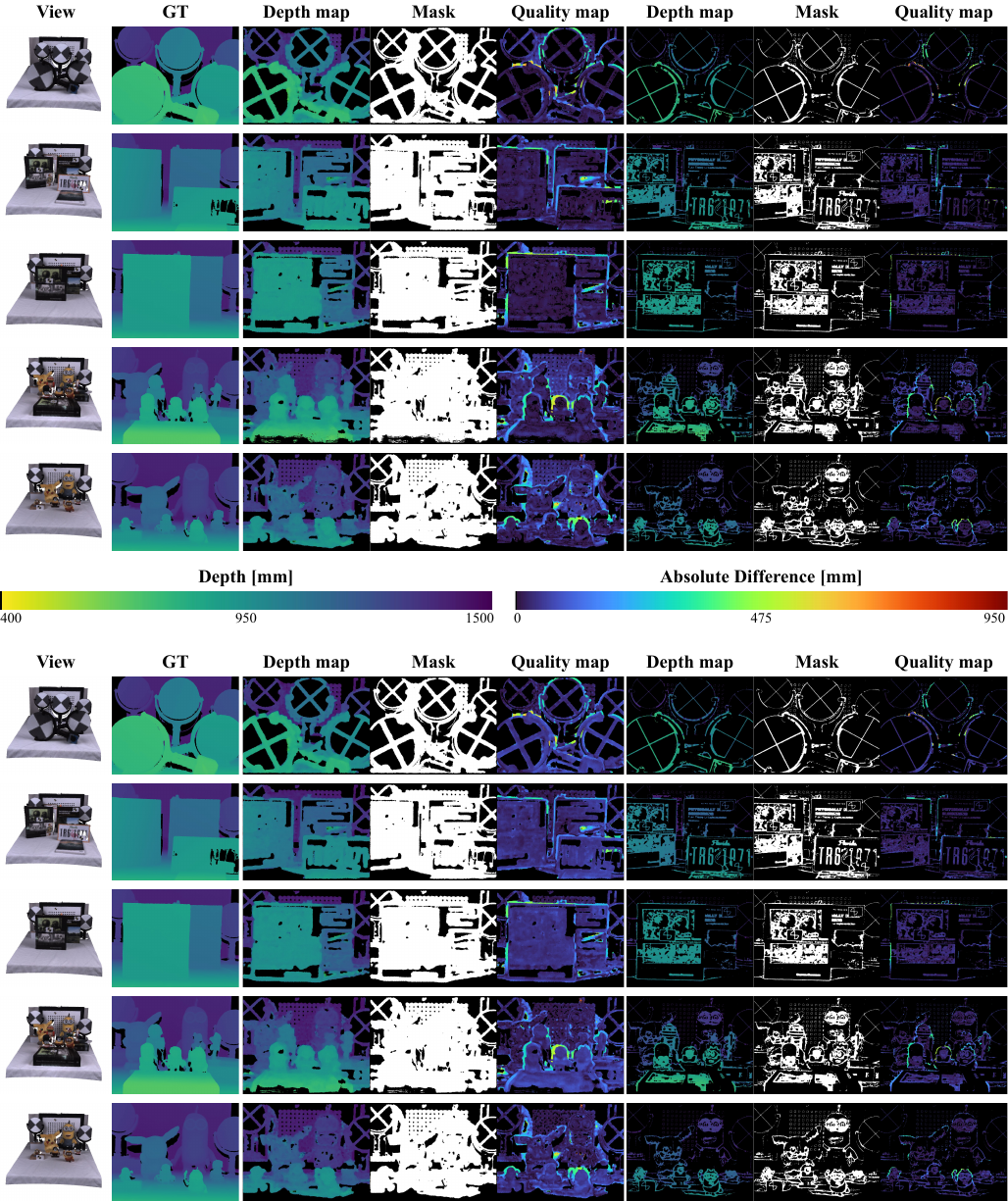}
	\caption{
		Snapshot view of the colored point cloud, along with the ground truth \ac{CSAD} are reported for each scene of the dataset \texttt{R12-ELP20}.
		\ac{CSAD}, mask and quality map representing the absolute difference (AD) error are illustrated for all variations using only the disparity (\texttt{D}) of our framework:
		\textit{top} is  the scaled (\texttt{S}) coarse (\texttt{C}) and refined variations (\texttt{R}), 
		\textit{bottom} is the unscaled (\texttt{U}) coarse (\texttt{C}) and refined variations (\texttt{R}).
		Please refer to the color version for better visualization.
	}\label{fig:3dscenedeptherrors-disp}
\end{figure*}

Snapshot of the colored point cloud, along with the ground truth \ac{CSAD} are reported for each scene in \autoref{fig:3dscenedeptherrors-blade} and \autoref{fig:3dscenedeptherrors-disp}.
Depth map, mask and quality map are illustrated for all variations using the relative blur (\texttt{B}) in \autoref{fig:3dscenedeptherrors-blade}, and for all variations using only the disparity (\texttt{D}) in \autoref{fig:3dscenedeptherrors-disp}.

\section{Source code and datasets redistribution}

The datasets and the source code are publicly available to the community.
Datasets can be downloaded from \url{https://github.com/comsee-research/plenoptic-datasets}, and more details can be found on this page.\\
We also developed an open-source C++ library for plenoptic camera named \texttt{libpleno} which is available at \url{https://github.com/comsee-research/libpleno}.\\
Along with this library, we developed a set of tools to calibrate a multi-focus plenoptic camera, named \texttt{COMPOTE} (standing for Calibration Of Multi-focus PlenOpTic camEra)  which is available at \url{https://github.com/comsee-research/compote}.\\
Finally, we developed a set of tools to estimate depth from raw images obtained by a multi-focus plenoptic camera, named \texttt{BLADE} (for BLur Aware Depth Estimation with a plenoptic camera) which is available at \url{https://github.com/comsee-research/blade}.

\begin{table*}[]
	\caption{
		Intrinsic parameters for datasets \texttt{R12-A,B,C} and \texttt{R12-E} obtained by our method \cite{Labussiere2020blur} (\texttt{BAP}), by the method of \citet{Noury2017b} (\texttt{NOUR}), and by \texttt{RxLive} software \cite{Heinze2016b} (\texttt{RTRX}). Blur proportionality coefficient and scaling coefficients are also reported for our method.
	}\label{tab:intrinsics}
	\centering\footnotesize
	\begin{tabu}{rl XXX XXX XXX X[3]}
		\toprule
		&&\multicolumn{3}{c}{\texttt{R12-A} ($\focusdist = 450~\si{\mm}$)} &\multicolumn{3}{c}{\texttt{R12-B} ($\focusdist = 1000~\si{\mm}$)} & \multicolumn{3}{c}{\texttt{R12-C} ($\focusdist = \infty$)} & \texttt{R12-E} ($\focusdist = 2133~\si{\mm}$)\\
		\cmidrule(r){3-5}\cmidrule(lr){6-8}\cmidrule(lr){9-11}\cmidrule(l){12-12}
		&& \texttt{BAP} & \texttt{NOUR} & \texttt{RTRX} & \texttt{BAP} & \texttt{NOUR} & \texttt{RTRX} & \texttt{BAP} & \texttt{NOUR} & \texttt{RTRX} & \texttt{BAP}\\
		\midrule
		\midrule
		$\Focal$& [\si{\mm}]&
		\nump{3}{49.885285868753407}  &\nump{3}{54.760830414143285} & \nump{3}{47.709} & %
		\nump{3}{50.010862604694566} &\nump{3}{51.177333513596956} &\nump{3}{50.8942} & %
		\nump{3}{50.099040536044029} &\nump{3}{51.644147026874265} &\nump{3}{51.5635} 
		& \nump{3}{50.119473840049658}\\ %
		$Q_1$& [$\times10^{-5}$]&
		\nump{2}{24.629793152694231}  &\nump{3}{6.1935769523508333} &- &  %
		\nump{3}{4.6612327268965204} &\nump{3}{1.650095415107347} &- &%
		\nump{2}{13.839048931748937} &\nump{3}{1.2915406728231928} &- 
		& -\nump{3}{6.823267739500167}\\
		$-Q_2$& [$\times10^{-6}$]&
		\nump{3}{3.0321615240649705}  &\nump{3}{0.80014173005714248}&- &  %
		\nump{3}{0.51566701947401107} &\nump{3}{0.2643386153887022}&- &%
		\nump{3}{2.72274849082607668} &\nump{3}{0.57639613295088305}&-
		& -\nump{3}{0.4075030150045897}\\
		$Q_3$& [$\times10^{-8}$]&
		\nump{3}{1.0948876088478752}  &\nump{3}{0.25176483785161299} &- &  %
		\nump{3}{0.15605904047468115} &\nump{3}{0.077672857411457887} &- &%
		\nump{3}{1.2598166788356634} &\nump{3}{0.18549080711917969} &-
		& -\nump{3}{0.046632026538200404}\\ %
		$P_1$& [$\times10^{-5}$]&
		-\nump{1}{11.07331475272971}  &-\nump{1}{18.05905135265203} &- &  %
		\nump{2}{12.844578250992004} &\nump{2}{11.266562984847101} &- &%
		\nump{2}{2.510763355364109} &\nump{2}{12.127867696914956} &-
		& \nump{3}{20.749099670679748}\\
		$-P_2$& [$\times10^{-5}$]&
		\nump{3}{3.5990205496035423}  &\nump{3}{5.1862878894614523} &- &  %
		\nump{2}{24.326834368558259} &\nump{2}{23.155561474048181} &- &%
		-\nump{3}{3.0723044374484207} &-\nump{3}{0.027303984846418686} &-
		& \nump{3}{11.12802972890747}\\
		\midrule
		$-Q_{{\text{-}1}}$& [$\times10^{-5}$]&
		\nump{2}{24.286960399023933}  &\nump{3}{6.195328231521894} &- &  %
		\nump{3}{4.6849029112134603} &\nump{3}{1.6968916342023989} &- &%
		\nump{2}{13.781154446130677} &\nump{3}{1.316444625218327} &- 
		& -\nump{3}{6.8530747139805667}\\
		$Q_{\text{-}2}$& [$\times10^{-6}$]&
		\nump{3}{2.9714668383971255}  &\nump{3}{0.81439516356456029}&- &  %
		\nump{3}{0.52820683947001753} &\nump{3}{0.27916224508984374}&- &%
		\nump{3}{2.7047099188535478} &\nump{3}{0.58914811295282422}&-
		& -\nump{3}{0.39435446259422714}\\
		$-Q_{\text{-}3}$& [$\times10^{-8}$]&
		\nump{3}{1.0662185708637543}  &\nump{3}{0.2577869990064612} &- &  %
		\nump{3}{0.15989121564334231} &\nump{3}{0.081791324078273949} &- &%
		\nump{3}{1.2461064824067977} &\nump{3}{0.18603523368734878} &-
		& -\nump{3}{0.037080422116950014}\\ %
		$-P_{\text{-}1}$& [$\times10^{-5}$]&
		-\nump{2}{10.892509120336487}  &-\nump{2}{18.000956166309254} &- &  %
		\nump{2}{12.85634701922286} &\nump{2}{11.315738281396624} &- &%
		\nump{2}{2.4973288978608905} &\nump{2}{12.171260719234855} &-
		& \nump{3}{21.030857515953071}\\
		$P_{\text{-}2}$& [$\times10^{-5}$]&
		\nump{3}{3.5396632437188866}  &\nump{3}{5.2178518402784784} &- &  %
		\nump{2}{24.467309667820016} &\nump{2}{23.3783090874790988} &- &%
		-\nump{3}{3.0672570039528202} &-\nump{3}{0.052752934478784024} &-
		& \nump{3}{11.324969500025701}\\
		\midrule
		\midrule
		$\Dist$& [\si{\mm}] &
		\nump{3}{56.859967077387466} &\nump{3}{62.341404683815092} & -&  %
		\nump{3}{52.140048198216938} &\nump{3}{53.2129351984062}&-& %
		\nump{3}{49.356326976170294} &\nump{3}{50.728182582087413} &-
		& \nump{3}{50.585393236494191}\\ %
		$-t_x$& [\si{\mm}] &
		\nump{2}{10.932145543311453} &\nump{3}{9.4798767949753611}&- & %
		\nump{2}{12.15142307076129} &\nump{2}{12.384668427457305} &- & %
		\nump{2}{12.527568800208069} &\nump{2}{13.242084890921969} &-
		& \nump{3}{12.875897977038225}\\
		$-t_y$& [\si{\mm}] &
		\nump{3}{7.9959730514471792} &\nump{3}{8.0874072752828283}&- &  %
		\nump{3}{6.1649549581760921} &\nump{3}{5.9650898062240589}&- & %
		\nump{3}{8.2366123244885525} &\nump{3}{7.4002652348034408}&- 
		& \nump{3}{6.6156396317421127}\\
		$-\theta_x$ &[\si{\micro\radian}] &
		\nump{1}{388.85519998996276} &\nump{1}{460.27738209856669}&- &
		\nump{1}{488.44314869479245} &\nump{1}{555.44031791339454}&- &
		\nump{1}{409.8388205523041} &\nump{1}{442.16617735663347}&- 
		& \nump{1}{441.6006983826283}\\
		$\theta_y$ &[\si{\micro\radian}] &
		\nump{1}{271.35879355952275} &\nump{1}{363.43239198422531}&- &
		\nump{1}{286.50476007926567} &\nump{1}{330.12322433414901}&- &
		\nump{1}{306.06459703077935} &\nump{1}{333.4245533371649}&- 
		& \nump{1}{289.22001690936499}\\
		$\theta_z$ &[\si{\micro\radian}] &
		\nump{1}{29.499664458583925} &\nump{1}{25.619255847084234} & \nump{1}{41.87424}&  
		\nump{1}{30.887328743121759} &\nump{1}{33.917291554850006}&\nump{1}{41.87424} &
		\nump{1}{33.928205745600323} &\nump{1}{39.916581369143224}&\nump{1}{36.6290384017}
		& \nump{1}{37.55805135840533}\\
		$\mlinterdist$ &[\si{\micro\m}] &
		\nump{2}{127.4552010932988} &\nump{2}{127.39671795575674}& \nump{2}{127.357684}&  %
		\nump{2}{127.47454872647138} &\nump{2}{127.40709817170356}& \nump{2}{127.357684}& %
		\nump{2}{127.45766281333226} &\nump{2}{127.411673369412349}& \nump{2}{127.357684}
		& \nump{2}{127.44587661565038}\\
		\midrule
		\midrule
		$\focal\mltype{1}$ &[\si{\micro\m}]&
		\nump{2}{582.66966865506242} &-&-&  %
		\nump{2}{566.38899095057971} &-&-& %
		\nump{2}{580.79528804213276} &-&-
		& \nump{2}{601.58440220844223}\\
		$\focal\mltype{2}$& [\si{\micro\m}]&
		\nump{2}{524.0249316658897} &-&-&  %
		\nump{2}{507.08733610974643}&-&-& %
		\nump{2}{515.57377489795142} &-&-
		& \nump{2}{562.18757048775958}\\
		$\focal\mltype{3}$ &[\si{\micro\m}]&
		\nump{2}{560.57361553341034} &-&-&  %
		\nump{2}{542.4685020176635} &-&-& %
		\nump{2}{552.83513789959582} &-&-
		&\nump{2}{583.53736817049551}\\
		\midrule
		\midrule
		$\principalpointx$& [\si{pix}]&
		$2078.3$ &$2343.4$ &- & %
		$1855.8$ & $1811.9$ &- & %
		$1786.6$ & $1654.9$ &-
		& $1722.5$\\
		$\principalpointy$ &[\si{pix}]&
		$1591.0$ &$1573.7$ &- & %
		$1926.2$ &$1962.2$ &- & %
		$1547.1$ & $1699.7$  &- 
		& $1843.6$\\ %
		$\distmlasensor$& [\si{\micro\m}]&
		\nump{2}{337.12571805114777}  &\nump{2}{391.89881708265517}&- &  %
		\nump{2}{326.71721233574402}  &\nump{2}{361.00537718834858}&- & %
		\nump{2}{330.31849375628042} & \nump{2}{357.81726862224161} &-
		& \nump{2}{340.86576314657435}\\ %
		\midrule
		\midrule
		$\kappa$ & - &
		\nump{3}{0.81335757753444282} &-&-&  %
		\nump{3}{0.77632365684589211} &-&-& %
		\nump{3}{0.74044876747258848} &-&-
		& \nump{3}{1.0201626906564538}\\
		\midrule
		$\gamma_2$ & [$\times10^{-5}$] &
		\nump{3}{79.708898900074192} &-&-&  %
		\nump{3}{23.017216184680102} &-&-& %
		\nump{3}{2.73552683795736498} &-&-
		& \nump{3}{4.6169969732325145}\\
		$\gamma_1$ & - &
		\nump{3}{0.62515220840000596} &-&-&  %
		\nump{3}{0.79609141585669219} &-&-& %
		\nump{3}{0.88338184462070446} &-&-
		& \nump{3}{0.91160914393108972}\\
		$\gamma_0$ & - &
		\nump{3}{31.51235516605901} &-&-&  %
		\nump{3}{10.912844125074178} &-&-& %
		\nump{3}{10.90983152629782} &-&-
		& \nump{3}{-2.004049745583496}\\
		\bottomrule
	\end{tabu}
\end{table*}

\end{document}